\documentclass[12pt,reqno]{article}
\usepackage{amsmath}
\usepackage{amsthm}
\usepackage{amsfonts}
\usepackage{amssymb}
\usepackage{epsfig}
\usepackage{graphics}
\usepackage{graphicx,indentfirst}
\numberwithin{equation}{section}

\def \Fig#1#2#3 {
\begin{figure}
\centering
\epsfxsize=#2cm \epsfbox{#1.eps}
\caption{#3}
\label{#1}
\end{figure}
}

\newcount\figno
\def\fig#1#2#3{
\par\begingroup\parindent=0pt\leftskip=1cm\rightskip=1cm\parindent=0pt
\baselineskip=15pt
\global\advance\figno by 1
\epsfxsize=#3
\centerline{\epsfbox{#2}}
\vskip 12pt
{\bf \small Figure \the\figno:} {\small #1}\par
\endgroup\par
}
\def\figlabel#1{\xdef#1{\the\figno
\mbox{ }}}
\def\encadremath#1{\vbox{\hrule\hbox{\vrule\kern8pt\vbox{\kern8pt
\hbox{$\displaystyle #1$}\kern8pt}
\kern8pt\vrule}\hrule}}

\def\L2{{\it Fun\/}\bigl(\text{\GL}\bigr)}

\newcommand{\chim}{\chi_-}
\newcommand{\chip}{\chi_+}
\newcommand{\chipm}{\chi_\pm}

\newcommand{\bm}{b_-}
\newcommand{\bp}{b_+}
\newcommand{\cm}{c_-}
\newcommand{\cp}{c_+}
\newcommand{\cpm}{c_\pm}

\newcommand{\del}{\partial}

\newcommand{\etam}{\eta_-}
\newcommand{\etap}{\eta_+}

\newcommand{\ad}{\text{ad}}

\def\gl{{\rm gl(1$|$1)}}
\def\GL{{\rm GL(1$|$1)}}

\newcommand{\sdet}{\text{sdet}}
\newcommand{\sign}{\text{sign}}
\newcommand{\tr}{\text{tr}}
\newcommand{\<}{\langle}
\renewcommand{\>}{\rangle}

\def\gl{{\rm gl(1$|$1)}}
\def\GL{{\rm GL(1$|$1)}}

\author{\\ Thomas Creutzig and Volker Schomerus\\[4mm]\small
  DESY Theory Group, DESY Hamburg\\\small
  Notkestrasse 85, D 22603 Hamburg, Germany}

\date{April 2008}

\begin{document}
\begin{titlepage}
\title{Boundary Correlators in Supergroup WZNW Models}
  \maketitle       \thispagestyle{empty}

\vskip1cm
\begin{abstract}
We investigate correlation functions for maximally symmetric
boundary conditions in the WZNW model on \GL. Special attention is
payed to volume filling branes. Generalizing earlier ideas for the
bulk sector, we set up a Kac-Wakimoto-like formalism for the
boundary model. This first order formalism is then used to
calculate bulk-boundary 2-point functions and the boundary 3-point
functions of the model. The note ends with a few comments on
correlation functions of atypical fields, point-like branes and
generalizations to other supergroups.
\end{abstract}

\vspace*{-14.9cm}\noindent
{\tt {yymm.nnnn}}\\
{\tt {DESY 08-042}}
%\bigskip\vfill
%\noindent
%\phantom{wwwx}{\small e-mail:}{\small\tt thomas.creutzig@desy.de,
%volker.schomerus@desy.de}
\end{titlepage}

\baselineskip=19pt \setcounter{equation}{0}
\tableofcontents

\section{Introduction}

Sigma models on supergroups and their cosets are an interesting
subject of current research. They occur in a number of very
different problems ranging from string theory to disordered
electron systems. In addition to such concrete applications,
conformal field theories with target space supersymmetry may also
be studied for their structural and mathematical properties. They
provide examples of non-unitary models, many of which have
vanishing or negative central charge. Moreover, their correlation
functions often possess logarithmic singularities. As shown in
\cite{Schomerus:2005bf}, both properties are intimately related to
features of the supergroup geometry.

The simplest non-trivial model to consider is the WZNW model on
the supergroup \GL. Studies of this field theory go back to the
work of Rozanski and Saleur
\cite{Rozansky:1992td,Rozansky:1992zt}. These early investigations
of the \GL\ WZNW model stimulated much further work on the
emerging topic of logarithmic conformal field theory (see e.g.\
\cite{Flohr:2001zs,Gaberdiel:2001tr} for a review). A few years
back, the \GL\ WZNW model was revisited in \cite{Schomerus:2005bf}
from a geometric rather than algebraic perspective. Based on the
harmonic analysis of the supergroup \GL, a proposal was formulated
for the exact spectrum of the field theory. Furthermore, efficient
computational tools were developed to calculate correlation
functions of tachyon vertex operators. Finally, the consistency of
the proposed spectrum was demonstrated explicitly.

The work \cite{Schomerus:2005bf} was restricted to the \GL\ WZNW
model on the sphere, i.e. neither boundaries nor higher genus
surfaces were included. Subsequent work \cite{Creutzig:2007jy}
extended part of the bulk analysis to the boundary sector. In
particular, the geometric interpretation of maximally symmetric
boundary conditions was unravelled. This led to several proposals
for the spectra of boundary operators in the corresponding
boundary conformal field theories. These were tested partially
through the so-called modular bootstrap. Correlation functions
with non-trivial insertions of bulk and boundary operators were
not computed in \cite{Creutzig:2007jy}. We are now aiming to close
this gap, at least for one type of boundary conditions.

There are several motivations to determine boundary correlation
functions in supergroup WZNW
models. To begin with, the conjectured boundary spectra in
\cite{Creutzig:2007jy} contained information that cannot be probed
through the modular bootstrap alone. In particular, certain boundary
correlation functions were predicted to contain logarithmic
singularities. Below we shall be able to verify such features of
the boundary conformal field theory. Moreover, 2-dimensional
boundary field theories are intimately related with quantization
theory (see e.g.
\cite{Schomerus:1999ug,Alekseev:1999bs,Alekseev:2002rj,Schomerus:2002dc}
and references therein). While the \GL\ WZNW model itself is a bit
too simple to accommodate for interesting supersymmetric
extensions of non-commutative geometry, the methods we shall
develop below possess generalizations to cases with a curved
bosonic base. The latter provide a much richer geometric
framework, with further links to representation theory of affine
algebras and the quantization of Lie superalgebras. Finally, let
us also mention possible applications to the study of branes and
open strings in superspaces, and in particular to $AdS$
backgrounds.

To be a bit more specific about the results we are going to
obtain, we recall from \cite{Creutzig:2007jy} that there are two
different families of maximally symmetric boundary conditions in
the \GL\ WZNW model. Geometrically, the first set consists of
D-branes that are point-like localized in the bosonic base. They
extend into both fermionic directions, unless they are placed
along very special lines in the base manifold. The second set of
boundary conditions contains a single object: a volume filling
brane that extends in all bosonic and fermionic directions. We
called this brane {\em twisted} because it is associated with the
only non-trivial gluing automorphism of the current algebra. In
\cite{Creutzig:2007jy}, some simple amplitudes for the point-like
D-branes have been computed. On the other hand, the methods of
\cite{Creutzig:2007jy} were not sufficient to obtain non-trivial
amplitudes for the volume filling brane.

In this work we shall extend some of the techniques from
\cite{Schomerus:2005bf} to compute correlation functions of bulk
and boundary operators for the volume filling brane. The main
results include explicit formulas (\ref{2pt},\ref{C0},\ref{C1})
for the bulk-boundary 2-point function and
(\ref{3t000}-\ref{3pt111}) for the boundary 3-point functions. The
information they contain is equivalent to the bulk-boundary and
the boundary operator product expansion, respectively. Our results
provide a complete solution of the boundary theory for the volume
filling brane. We shall also determine a non-trivial annulus
amplitude.

In order to obtain these results we set up a first order formalism
for the volume filling brane. It is obtained by adding an appropriate
square root of the bulk interaction term along the boundary of the
world-sheet. As in other theories containing fermions, taking the
square root forces us to introduce an auxiliary fermion along the
boundary. All this will be explained in great detail in section 2.
A perturbative expansion for correlators of the boundary conformal
field theory is set up in section 3. It is employed in Section 4
to solve explicitly the boundary \GL\ WZNW model with twisted
boundary conditions. Section 5 contains an alternative approach to
computing amplitudes that involve only special (atypical)
fields/states of the theory. It is used to prove that the \GL\ WZNW
contains a special subsector whose correlation functions are
independent of the level $k$. The second approach is finally
employed to compute a particular annulus amplitude for the volume
filling brane. The latter provides a nice test for the boundary
state that was proposed in \cite{Creutzig:2007jy}. We conclude with
a list of open problems, mostly related to the point-like
branes for \GL\ and extensions to higher supergroups.

\section{Volume filling brane: The classical action}

Our aim in this section is to discuss the classical
description of volume filling branes in the \GL\ WZNW model. To
begin with, we spell out the standard action of the WZNW model
with so-called twisted boundary conditions. Their geometric
interpretation as volume filling branes with a non-zero B-field is
recalled briefly. In order to set up a successful computation
scheme for the quantum theory later on, we shall need a different
formulation of the theory. As in the bulk theory, computations of
correlations functions require a Kac-Wakimoto like representation
of the model \cite{Schomerus:2005bf}. Finding such a first order
formalism for the boundary theory is not entirely straightforward.
We shall see that it requires introducing an additional fermionic
boundary field.

\subsection{The boundary WZNW model}

Following our earlier work on WZNW models for type I supergroups,
we parametrize the supergroup \GL\ through a Gauss-like
decomposition of the form
$$ g \ = \ e^{i \eta_- \psi^-} \, e^{ixE + iyN} \, e^{i \eta_+
\psi^+} $$ where $E,N$ and $\psi^\pm$ denote bosonic and fermionic
generators of \gl, respectively. In the WZNW model, the two even
coordinates $x,y$ become bosonic fields $X,Y$ and similarly, two
fermionic fields $c_\pm$ come with the odd coordinates $\eta_\pm$.
Let us now consider a boundary WZNW model with the action
\begin{equation}\label{eq:SWZW}
    \begin{split}
S_{\text{WZNW}}(X,Y,c_\pm)\ =\  &-\frac{k}{4\pi i}\int_\Sigma
d^2z\
  \left( \del X\bar{\del}Y+\del
  Y\bar{\del}X+2e^{iY}\del\cp\bar{\del}\cm\right)\,  + \\[2mm]
        & +\frac{k}{8\pi i}\int du \ e^{iY}(\cp+\cm)\del_u(\cp+\cm)\
        ,\\
    \end{split}
\end{equation}
where $u$ parametrizes the boundary of the upper half plane.
Variation of the action leads to the usual bulk equations of
motion along with the following set of boundary conditions\\[0mm]
\begin{equation}\label{eq:bdyeom}
    \begin{split}
      \del_v Y\ = \ 0 \ \ \ \ & , \ \ \ \
         2 \del_v X \ = \ e^{iY}(\cp+\cm)\,  \del_u(\cp+\cm)\ ,
         \\[3mm]
      \pm 2  \del_v c_\pm\, & = \ 2 i \del_u c_\mp - (\cm+\cp)\,
       \del_u Y \ . \\[3mm]
    \end{split}
\end{equation}
Here, we have used the derivatives $\del_u = \del + \bar{\del}$
and $\del_v = i (\del - \bar{\del})$ along and perpendicular to
the boundary. The equations \eqref{eq:bdyeom} imply Neumann
boundary conditions for all four fields of our theory, i.e.\ we
are dealing with a volume filling brane. Since the normal
derivatives of the fields $X$ and $c_\pm$ do not vanish, our brane
comes equipped with a B-field. A more detailed discussion of the
brane's geometry can be found in our recent paper
\cite{Creutzig:2007jy}.

In order to see that our boundary conditions preserve the full
chiral symmetry, we recall that the holomorphic currents of the
\GL\ WZNW model take the form
\begin{equation*}
    \begin{split}
    J^E\, =\ ik\del Y \ \ \ \ \ \ \ \ \ & , \ \ \ \
    J^N\, =\ ik\del X -k\cm \del\cp \, e^{iY} \ , \\[2mm]
    J^-\, =\ -ke^{iY}\del\cp\  \ \  & , \ \ \ \
    J^+ \, = \ k\del\cm +ik\cm \del Y \ , \\[1mm]
\end{split}
\end{equation*}
and similarly for the anti-holomorphic currents,
\begin{equation*}
    \begin{split}
        \bar{J}^E \, =\ -ik\bar{\del} Y \ \ \ \ \ \ \ \  & , \ \ \ \
        \bar{J}^N\, =\ -ik\bar{\del} X +k \bar{\del}\cm\, \cp\, e^{iY} \ ,
        \\[2mm]
            \bar{J}^+\, =\ -ke^{iY}\bar{\del}\cm \  \
            \ \ & , \ \ \ \
        \bar{J}^-\, =\ k\bar{\del}\cp+ik\cp\bar{\del} Y \ . \ \
        \\[1mm]
\end{split}
\end{equation*}
If we plug the boundary conditions \eqref{eq:bdyeom} into these
expressions for chiral currents, we obtain the gluing condition
$ J^X(z) = \Omega \bar J^X(\bar z)$ for $X=E,N,\pm$ and all along
the boundary at $z= \bar z$. Here, the relevant gluing automorphism
$\Omega$ is obtained by lifting the automorphism
\begin{equation}
    \label{eq:twisted}
    \Omega(E)\, = \,  - E, \ \ \Omega(N)\, =\,  - N
    , \ \ \Omega(\psi^+) \, =\,  - \psi^-, \ \ \Omega(\psi^-)\,
   = \, \psi^+\  
\end{equation}
from the finite dimensional superalgebra \gl\ to the full affine
symmetry. In \cite{Creutzig:2007jy} we called these gluing
conditions {\em twisted} and showed that there is a unique brane
corresponding to this particular choice of $\Omega$.

\subsection{First order formulation}

Computations of bulk and boundary correlators in the presence of
twisted D-branes shall be performed in a first order formalism. In
the bulk, it is well-known how this works \cite{Schomerus:2005bf}.
There, the bulk action is built of a free field theory involving
two additional fermionic auxiliary fields $b_\pm$ of weight
$\Delta(b_\pm) = 1$ along with the original fields $X,Y$ and
$c_\pm$,
\begin{equation}
    \begin{split}
        S_{0;\text{cl}}^{\text{bulk}}[X,Y,c_\pm,b_\pm] \ = \
        &-\frac{k}{4\pi i}\int_\Sigma d^2z\ \left( \del X\bar{\del}Y+
        \del Y\bar{\del}X\right)
        \\[2mm]
         &-\frac{1}{2\pi i}\int_\Sigma d^2z\
         \left(\bp\del\cp+\bm\bar{\del}\cm\right)  \ . \\
        \end{split}
\end{equation}
We placed a subscript `cl' on the actin to distinguish it from the
action we shall use in our path integral computations later on. If
the following bulk marginal interaction term is added to the free
field theory,
\begin{equation}
    \label{eq:Sbulk}
    S^{\text{bulk}}_{\text int}[X,Y,c_\pm,b_\pm] \ = \
    -\frac{1}{2k\pi i}\int_\Sigma d^2z\ e^{-iY}\bm\bp \ 
\end{equation}
the equations of motion for $b_\pm$ read $b_- = k \del c_+ \exp
iY$ and $b_+ = - k \bar \del c_- \exp iY$ so that we recover the
bulk WZNW-model upon insertion into the first order action. In
extending this treatment to the boundary sector, we are tempted to
add the ``square root'' of the bulk interaction as a boundary
term. This is indeed what happens for the closely related $AdS_2$
branes in $AdS_3$ \cite{Fateev:2007}. Here, however, it cannot
possibly be the right answer, at least not without a proper notion
of what we mean by taking the square root. In fact, the naive
square root of $\bm\bp\exp(-iY)$ is something like
$b_\pm\exp(-iY/2)$, i.e.\ a fermionic operator. It makes no sense
to add such an object to the bulk theory. In order to take a
bosonic square root of the bulk interaction, we introduce a new
fermionic boundary field $C$ of weight $\Delta(C) = 0$ and add
the following terms to the bulk theory,
\begin{eqnarray}
    S^{\text{bdy}}_0[X,Y,c_\pm,b_\pm,C] & = &
    \frac{1}{8\pi i}\int du \ \left( kC\del_uC+4(\cp+\cm)\bp\right)  \\[2mm]
    S^{\text{bdy}}_{\text{int}}[X,Y,c_\pm,b_\pm,C]
     & = & -\frac{1}{2\pi i}\int du \ e^{-iY/2}\bp C \ .
\end{eqnarray}
The idea to involve an additional fermionic boundary field in the
action of supersymmetric brane configurations is not new. It was
initially proposed in \cite{Warner:1995ay} and has been put to use
more recently \cite{Kapustin:2003ga,Brunner:2003dc} in the context
of matrix factorizations. Our boundary action resembles the one
Hosomichi employed to treat branes in $N=2$ super Liouville theory
\cite{Hosomichi:2004ph}. The full \gl\ boundary theory now takes
the form
\begin{equation}
    S[X,Y,\cpm,b_\pm,C] \ = \ S_{0,\text{cl}}^{\text{bulk}} +
      S_0^{\text{bdy}} + S^{\text{bulk}}_{\text{int}}+
       S^{\text{bdy}}_{\text{int}} \ = \ S_{0,\text{cl}} + S_{\text{int}}
\end{equation}
where
\begin{equation}
 \begin{split}
         S_{0,\text{cl}} \ = & \ -\frac{k}{4\pi i}\int_\Sigma d^2z\ \left(\del X\bar{\del}Y+
             \del Y\bar{\del}X \right)   \\[2mm]
         & - \frac{1}{2\pi i}\int_\Sigma d^2z\
         \left(\cp\del\bp+\cm\bar{\del}\bm\right)
         +\frac{1}{8\pi i}\int du \ kC\del_uC \ ,\\[3mm]
     S_{\text{int}} \ =  & \ -\frac{1}{2k\pi i}\int_\Sigma d^2z\ e^{-iY}\bm\bp \
     - \frac{1}{2\pi i}\int du \ e^{-iY/2}\bp C\ \ .  \\
     \end{split}
\end{equation}
Here, we have performed a partial integration on the kinetic term
for the bc-system, thereby absorbing the contribution $b_+(c_- +
c_+)$ from the boundary action. This is similar to the
case of $AdS_2$ branes in $AdS_3$ \cite{Fateev:2007}. In order to
complete the description of the classical action, we add the
following Dirichlet boundary condition for the fields $b_\pm$,
\begin{equation}
b_+(z) + b_-(\bar z) \ = \ 0 \ \ \ \mbox{ for } \ \ z \ = \ \bar z
\ .
\end{equation}
If the action is varied with this boundary condition, we recover
the boundary equations of motion \eqref{eq:bdyeom}. More
precisely, we obtain four equations among boundary fields. Two of
these can be used to determine the boundary fields $C$ and $b_+ =
- b_-$ through $X,Y$ and $c_\pm$,
\begin{equation}\label{Cbpm}
 C \ = \ \, e^{iY/2} \, (\cp + \cm) \ \ \ , \ \ \ \pm2 b_\pm \ = \
 k\, e^{iY/2} \partial_u C \ .
\end{equation}
The four equations among boundary fields along with the bulk
equations motion for $b_\pm$ imply the eqs. \eqref{eq:bdyeom}. We
leave the details of this simple computation to the reader.

We have now set up a first order formalism for the twisted brane
on GL(1$|$1). Let us stress again that is was necessary to
introduce an additional fermionic field $C$ on the boundary of the
world-sheet. Above we have motivated this new degree of freedom by
our desire to take a bosonic square root of the bulk interactions.
But there is another, more geometric, way to argue for the
additional field $C$. We mentioned before that the
first order formalism for the GL(1$|$1) WZNW model is very similar
to that for the Euclidean $AdS_3$, only that the bosonic
coordinates $\gamma, \bar \gamma$ of the latter are replaced by
fermionic ones. The first order formalism for $AdS_2$ branes in
$AdS_3$ was set up in \cite{Fateev:2007} and it describes a brane
that is localized along a 1-dimensional subspace of the
$\gamma\bar \gamma$ plane. Correspondingly, only a single $\gamma$
zero mode remains after imposing the boundary conditions. The
brane on GL(1$|$1) we are attempting to describe, however, is
volume filling and therefore it extends in both fermionic
directions. Therefore, we need two independent fermionic zero
modes. These are provided by the zero modes of the three fields
$c_\pm$ and $C$. Note that these fields are related by equation
\eqref{Cbpm}.

\section{Volume filling branes: The quantum theory}

Our next step is to develop a computational scheme for correlation
functions in the boundary WZNW model with twisted boundary
conditions. We shall use the first order formulation of section
2.2 as our starting point and consider the full WZNW model as a
deformation of a free field theory involving the fields
$X,Y,c_\pm,b_\pm$ and the fermionic boundary field $C$. This free
field theory will be described in more detail in the first
subsection. The definition of vertex operators and their
correlation functions in the WZNW model is the subject of
subsection 3.2.

\subsection{The free theory and its correlation functions}

Our strategy is to employ the first order formulation we set up in
the previous section. In order to do so, we have to add a few
comments on the measures we are using in the path integral
treatment. To begin with, the supergroup invariant measure of the
WZNW model is given by
\begin{equation} \label{WZNWmeasure}
    d\mu_{\text{WZNW}} \ \sim \ \, \mathcal{D}X\mathcal{D}Y
      \mathcal{D}(e^{iY/2} \cm)\mathcal{D}(e^{iY/2}\cp)\ .
\end{equation}
This gets multiplied with ${\cal D}b_+ {\cal D}b_- {\cal D} C$
when we pass to the first order formalism. But in the following we
would like to employ the standard free field measure
$$ d \mu_{\text{free}}\ \sim \ \mathcal{D}X\mathcal{D}Y
      \mathcal{D}\cm\mathcal{D}\cp\ .
      $$
The two measures are related by a Jacobian of the form (see
e.g.\ \cite{Gerasimov:1990} for similar computations)
\begin{equation}
 \begin{split}\label{measure}
    d\mu_{\text{WZNW}} \ &=  \  \left(\sdet(
  G^{ab} e^{iY}{\del_a}e^{-iY}\del_b)\right)^{-1} \ d\mu_{\text{free}} \\[3mm]
  \ &=  \ e^{\frac{1}{8\pi}\int dudv\ \sqrt{G}
   (-G^{ab}\del_a\  Y{\del}_bY+i\mathcal{R}Y) +\frac{1}{8\pi}\int du\
   i\sqrt{G}\mathcal{K}Y } \ d\mu_{\text{free}}  .\\
\end{split}
\end{equation}
Here, $G_{ab}$ is the metric on the world-sheet, ${\cal R} =
\del_a \del^a \log G $ and ${\cal K} = \frac{1}{2i}\partial_v \log G $
are its Gaussian and geodesic curvature, respectively. These two
quantities feature in the Gauss-Bonnet theorem for surfaces with
boundary,
\begin{equation}
 \frac{1}{4\pi}\int_\Sigma dudv\ \sqrt{G} \mathcal{R}+\frac{1}{4\pi}\int du\
 \sqrt{G}\mathcal{K}  \ = \ \chi(\Sigma) \ = \ 1 \ ,
\end{equation}
where $\chi(\Sigma) = 1$ is the Euler characteristic of the disc.
We can now pass to the upper half plane again where all curvature
is concentrated at infinity. The effect of the curvature terms in
the WZNW measure is to insert a background charge $Q_Y =
\chi(\Sigma)/2 = 1/2 $ for the field $Y$ at infinity. In addition,
the measure \eqref{measure} also contains a term that is quadratic
in $Y$. We simply add this to the free part of our action, i.e.\
we define
\begin{equation}
  \begin{split} \label{qS0}
   S_0 \ = &\  -\frac{1}{4\pi i}\int_\Sigma d^2z\ \left( k\, \del X\bar{\del}Y+
             k \, \del  Y\bar{\del}X
         -  \del Y\bar{\del}Y\right)  \\[2mm]
         & - \frac{1}{2\pi i}\int_\Sigma d^2z\
         \left(\cp\del\bp+\cm\bar{\del}\bm\right)
         +\frac{1}{8\pi i}\int du \ kC\del_uC \ ,
  \end{split}
\end{equation}
Note, that the new term in the actions modifies the formula for
the current $J^N$ by adding an additional $\del Y$ and similarly
for the anti-holomorphic partner.

In our path integral we now integrate with the free field theory
measure $d \mu_{\text{free}}$ over all fields subject to the
boundary condition $b_+ + b_- = 0$. Configurations for the other
fields are not constrained in the path integral. In the free
quantum field theory, they satisfy the linear (``Neumann'')
boundary conditions
\begin{equation}
    \begin{split}
       \del_v Y \ = & \ 0 \ \ \ \ , \ \ \ \ \ \ \ \,
       \del_v  X \ = \ 0  \ , \\[2mm]
       \del_u C \ = & \ 0 \ \ \ \ , \ \ \
       \cp+\cm\ = \ 0 \ \ . \\
       \end{split}
\end{equation}
These equations are satisfied in all correlation functions or,
equivalently, as operator equations on the state space of the free
field theory. Note that, according to the last equation, the zero
modes of $c_+$ and $c_-$ coincide in our free boundary theory. The
necessary second fermionic zero mode is exactly what is provided
by the field $C$.

Arbitrary correlation functions in the free field theory can now
easily be computed with the help of Wick's theorem. All we need to
use is the following list of operator product expansions
\begin{equation}
    \begin{split}
         X(z,\bar{z})Y(z,\bar{z})  \ \sim \ &
          \frac{1}{k}\ln|z-w|^2+\frac{1}{k}\ln|z-\bar{w}|^2
          \\[2mm]
         \cm(z)b_-(w)  \ \sim \ \ \frac{1}{w-z} &  \ \
        \ \
         \cp(\bar{z})b_+(\bar{w})  \ \sim \
         \frac{1}{\bar{w}-\bar{z}}\\[2mm]
         \cm(z)b_+(\bar{w}) \ \sim \
         \frac{1}{z-\bar{w}} & \ \ \ \
         \cp(\bar{z})b_-(w)  \ \sim  \
         \frac{1}{\bar{z}-w}\\[2mm]
         C(v)C(u) \ &\ \sim \  \frac{2\pi i}{k}\sign(v-u)\ \ \ .
    \end{split}
\end{equation}
Let us remark that a non-vanishing correlation function in the
free field theory requires that the fields $c$ outnumber the
insertions of $b$ by one. Furthermore, $C$ must be inserted an odd
number of times. We also recall that there is a non-vanishing
background charge $Q_Y = 1/2$ for the field $Y$. On the disk, the
corresponding U(1) charges of all tachyon vertex operators must
add up to $Q_Y \chi(\Sigma) = 1/2$ in order for the correlator to
be non-zero. These rules imply that the 1-point function of the
bulk identity field vanishes. In order to normalize the vacuum
expectation value, we require that
\begin{equation}
    \langle \, \left(\cm(z)-\cp(\bar{z})\right) \, C(u) \,
     e^{ieX(z,\bar z) + inY(z,\bar{z})}\, \rangle_0 \ = \
      \delta(e) \delta(n-1/2)  \ .
\end{equation}
Note that the product of fields in brackets is the simplest
expression that meets all our requirements: The U(1)$_Y$ charge of
the tachyon vertex operators is $m = 1/2$, we inserted one $c_\pm$
and no field $b_\pm$ and multiplied with a single $C$ in order to
make the total insertion bosonic again.

\subsection{Correlation functions in boundary WZNW model}

Now that we have learned how to perform computations in the free
field theory described by the action \eqref{qS0}, we would like
to add our interaction term
\begin{equation}\label{Sint}
    S_{\text{int}} \ = \
    -\frac{1}{2k\pi i}\int_\Sigma d^2z\ e^{-iY}\bm\bp \ -
    \frac{1}{2\pi i}\int du \ e^{-iY/2}\bp C \ .
\end{equation}
The idea is to calculate correlators of the full boundary
WZNW model perturbatively, i.e.\ by expanding the exponential of
the interaction in a power series. Even though there is a priori
an infinite number of terms to be considered, only finitely many
contribute to our perturbative expansion. This is very similar to
what has been observed in the bulk model \cite{Schomerus:2005bf}.

Before we can spell out precise formulas for the quantities we
want to compute, we need to explain how to associate free field
theory vertex operators to the fields of the interacting WZNW
model. The latter are in one-to-one correspondence with functions
on the supergroup GL(1$|$1) and they may be characterized by
their behavior with respect to global gl(1$|$1) transformations.
We shall first recall from \cite{Schomerus:2005bf} how this works
for bulk fields.

Let us begin by collecting a few basic facts about the space of
functions on the supergroup \GL\ \cite{Schomerus:2005bf}. As for
any other group or supergroup, $\L2$ carries two graded-commuting
actions of the Lie superalgebra gl(1$|$1). These are generated by
the following right and left invariant vector
fields\\[-2mm]
\begin{equation}
\begin{array}{llll}
 R_E = i\del_x \ , & R_N  = i\del_y+\etam\del_- \ ,
& R_+ = -e^{-iy}\del_+-i\etam\del_x \ ,  & R_- = -\del_-\ , \\[4mm]
 L_E = -i\del_x \ ,  & L_N = -i\del_y-\etap\del_+ \ ,
& L_- = e^{-iy}\del_- -i\etap\del_x \ ,  & L_+ = \del_+\ .\\[3mm]
\end{array}
\end{equation}
A typical irreducible multiplet for \gl\ is 2-dimensional. Hence,
typical irreducible multiplets of the combined left and right
action are spanned by four functions in the supergroup. As in
\cite{Schomerus:2005bf} we shall combine these functions into a
$2\times 2$ matrix of the form
\begin{equation}
    \begin{split}\label{phi}
        \varphi_{\<-e,-n+1\>}\ = \ e^{iex+iny}
        \left( \begin{array}{cc}
            1& \etam    \\
                 \etap &e^{-1}e^{-iy}+\etap\etam \\
                        \end{array}
                           \right)
        \end{split}
\end{equation}
The rows span the typical irreducibles $\<-e,-n+1\>$ of the right
regular action. Columns transform in the  representations
$\<e,n\>$ of the left regular action. Note that $\varphi_{\<
e,n\>}$ is only well defined for $e\neq0$, i.e.\ in the typical
sector of the minisuperspace theory.

Following \cite{Schomerus:2005bf}, the bulk vertex operators in
the free field theory are modelled after the matrices
$\varphi_{\<e,n\>}$. More precisely, let us introduce typical bulk
operators through
\begin{equation}
    \begin{split}\label{vertV}
        V_{\<-e,-n+1\>}(z,\bar{z})\ = \ e^{ieX+inY}
        \left( \begin{array}{cc}
                     1& \cm  \\
                 \cp & \cp\cm \\
                        \end{array}
                           \right)
        \end{split}
\end{equation}
Since the weight of the fermionic fields $c_\pm$ vanishes,
all four fields in this matrix possess the same conformal
dimension,
\begin{equation}
    \Delta_{(e,n)} \ = \ \frac{e}{2k}(2n-1+\frac{e}{k})\ .
\end{equation}
Note that one of the terms in the lower left corner of the
minisuperspace matrix $\varphi_{\<e,n\>}$ has no analogue on the
vertex operator $V_{\<-e,-n+1\>}$. We consider this term as
`subleading'. It is reconstructed when we build correlation
functions of the interacting WZNW model (see
\cite{Schomerus:2005bf} and \cite{Gotz:2006qp} for more details).
\smallskip

Let us now repeat the previous analysis for the boundary fields.
Since our twisted brane is volume filling, the relevant space of
minisuperspace wave functions is again the space $\L2$ of all
functions on the supergroup \GL . But this time, it comes equipped
with a different action of the Lie superalgebra \gl. In fact,
minisuperspace wave functions as well as boundary vertex operators
are now distinguished by their transformation under a single twisted
adjoint action $\ad^{\Omega}_X = R_X + L^\Omega_X$ of \GL\ on
$\L2$. Explicitly, the generators of \gl\ transformations are
given by
\begin{equation}
    \begin{split}\label{eq:twad}
      \ad_E^\Omega\ &=\ 2i\del_x \qquad\qquad , \ \
      \ad_N^\Omega\ =\ 2i\del_y+\etap\del_++\etam\del_- \ , \\
      \ad_-^\Omega\ &=\ \del_+-\del_- \qquad , \ \
      \ad_+^\Omega\ =\ -e^{-iy}(\del_-+\del_+)+i(\etap-\etam)\del_x \ \ . \\
    \end{split}
\end{equation}
Under the twisted adjoint action of \gl\ on $\L2$, each typical
multiplet appears with two-fold multiplicity
\cite{Creutzig:2007jy}. Once more, we propose to assemble the
corresponding four functions into a $2\times 2$ matrix of the form
\begin{equation}
    \begin{split} \label{eq:bdrmat}
        \psi_{\<-2e,-2n+1\>}\ = \ e^{iex+iny}
        \left( \begin{array}{cc}
                             1& \etap-\etam \\
                 \eta &2e^{-1}e^{-iy/2}+(\etap-\etam)\eta \\
                        \end{array}
                           \right)
        \end{split}
\end{equation}
where we introduced the shorthand $\eta=e^{iy/2}(\etam+\etap)$.
The reader is invited to check that the two rows of this matrix
each span the 2-dimensional typical irreducible $\<-2e,-2n+1\>$
under the twisted adjoint action \eqref{eq:twad} of the
superalgebra \gl.

Boundary vertex operators are modelled after the matrices $
\psi_{\<-2e,-2n+1\>}$ more or less in the same way as in the case
of bulk fields,
\begin{equation}
    \begin{split}\label{vertU}
        U_{\<-2e,-2n+1\>}(u)\ = \ e^{ieX+inY}
        \left( \begin{array}{cc}
                     1& \cp-\cm   \\
                 C & (\cp-\cm)C \\
                        \end{array}
                           \right)\ \ .
        \end{split}
\end{equation}
Again, we dropped the $y$-dependent term in the lower right corner
of the matrix \eqref{eq:bdrmat}. Eventually, we will see how this
term is recovered in boundary correlation functions. The main new
aspect of the prescription \eqref{vertU}, however, concerns the
appearance of the fermionic boundary field $C$ that we inserted in
place of the function $\eta$. This substitution is motivated by
the classical equation of motion \eqref{Cbpm}.
\smallskip

After this preparation we are able to spell out how correlation
functions of bulk and boundary fields can be computed for the
interacting WZNW model. More precisely, we define,
\begin{equation}
    \begin{split} \label{corrdef}
  \left\langle \prod_{\nu=1}^m\Phi_{\langle
        e_\nu,n_\nu\rangle}
         (z_\nu,\bar{z}_\nu)\right.&\left.\prod_{\mu=1}^{m'}
          \Psi_{\langle e_\mu,n_\mu\rangle}(u_\mu)
          \right\rangle \ =
          \\[2mm]
        & \sum_{s=0}^\infty\, \frac{(-1)^s}{s!}\ \left\langle
        \,  \left(S_{\text{int}}\right)^s \
        \prod_{\nu=1}^m V_{\langle e_\nu,n_\nu\rangle}(z_\nu,\bar{z}_\nu)
        \prod_{\mu=1}^{m'}
U_{\langle e_\mu,n_\mu\rangle}(u_\mu)\right\rangle_0 \ . \\
    \end{split}
\end{equation}
Here, $S_{\text{int}}$ is the interaction \eqref{Sint} and all
correlation functions on the right side are to be computed in the
free field theory \eqref{qS0}. The relevant vertex operators $V$
and $U$ were introduced in equations \eqref{vertV} and
\eqref{vertU} above. For later use we also note that bosonic
correlators can be determined by means of the following standard
formula,
\begin{eqnarray} \nonumber
        & &\hspace*{-5mm} \left\langle\prod_{\nu=1}^m V_{(e_\nu,n_\nu)}(z_\nu,\bar{z}_\nu)
        \prod_{\lambda=1}^{m'} V_{(e_\lambda,n_\lambda)}(u_\lambda)\right\rangle
          = \delta({\textstyle \sum}_{\nu=1}^mn_\nu+{\textstyle \sum}_{\lambda=1}^{m'}
          n_\lambda+{\textstyle{\frac12}})
         \delta({\textstyle \sum}_{\nu=1}^me_\nu+{\textstyle \sum}_{\lambda=1}^{m'}e_\lambda)\\[3mm]
        & & \ \ \ \
         \times \ \prod_{\nu>\mu}|z_\nu-z_\mu|^{-2\alpha_{\nu\mu}}\prod_{\nu>\mu}
        |z_\nu-\bar{z}_\mu|^{-2\alpha_{\nu\mu}}
         \prod_{\nu,\lambda}|z_\nu-u_\lambda|^{-4\alpha_{\nu\lambda}}
         \prod_{\lambda>\kappa}|u_\lambda-u_\kappa|^{-4\alpha_{\kappa\lambda}}
            \\[5mm]
          &  & \quad  \text{where} \qquad \qquad \qquad \alpha_{\nu\mu}\ =\
        -n_\nu\frac{e_\mu}{k}-n_\mu\frac{e_\nu}{k}-\frac{e_\nu
        e_\mu}{k^2}
\nonumber
\end{eqnarray}
and $V_{(e_\nu,n_\nu)}= \exp(ieX+inY)$ are bosonic vertex
operators. As in the bulk theory it is easy to see that the all
expansions \eqref{corrdef} truncate after a finite number of
terms. In fact, the inserted bulk and boundary vertex operators on
the right hand side of eq.\ \eqref{corrdef} contain at most $2m +
m'$ fermionic fields $c_\pm$. Since each interaction term from
$S_{\text{int}}$ contributes at least one insertion of $b_\pm$, we
conclude that terms with $s \geq 2m + m'$ vanish.

\section{Solution of the boundary WZNW model}

A boundary conformal field theory is uniquely characterized by the
bulk-boundary and the boundary operator product expansions. We
shall now employ the perturbative calculational scheme we developed
in the previous section in order to determine these data. After
a short warm-up with the discussion of bulk 1-point functions, we
determine the bulk-boundary 2-point function in the second
subsection. The 3-point function of boundary fields is addressed
in subsection 4.3.

\subsection{Bulk 1-point function}

The bulk 1-point function is the simplest non-vanishing quantity
in a boundary conformal field theory. It contains the same
information as the boundary state. For volume filling branes, the
boundary state was determined in our previous work
\cite{Creutzig:2007jy}. Our first aim now is to reproduce our old
result through our new perturbative expansion.

The 1-point function of a typical bulk field $\Phi_{\<e,n\>}$ is
computed by inserting a single vertex operator \eqref{vertV} into
the expansion \eqref{corrdef}. Since bulk vertex operators contain
at most two fields $c$, the only non-zero terms can come from
$s=0,1$. The term with $s=0$ contains no insertion of the
interaction and it vanishes identically. So, let us see what
happens for $s=1$. In this case, only the insertion of the
boundary interaction can contribute. The results is
\begin{eqnarray*}
    & & \langle \Phi_{\langle e,n\rangle}(z,\bar{z})\rangle \ = \
    \frac{i}{2\pi} \int du\ \langle e^{-iY(u)/2}b_+(u)C(u)
    V_{\<e,n\>}(z,\bar z)\rangle\\[2mm]
   &  &  \ \ \ \ = \ E^1_1 \delta(e) \delta(n-1) \frac{1}{4\pi i}
    \int du\ \left(\frac{1}{u-\bar z} - \frac{1}{u-z}\right)
    \ = \  \int d\mu\ \varphi_{\langle e,n\rangle} \ .
\end{eqnarray*}
Here, $E^1_1$ is the elementary matrix which has zeroes everywhere
except in the lower right corner. Note that the only field with
non-vanishing 1-point function has conformal weight $\Delta =0$.
Hence, there is no dependence on the insertion point $(z,\bar z)$.
In the last line we have expressed the numerical result as an
integral of the matrix valued function \eqref{phi} over the
supergroup \GL. The integration is performed with the Haar
measure
\begin{equation} d\mu \ = \ 2^{-1} e^{-iy}dxdyd\etap d\etam\ \ .
\end{equation}
Since the Haar measure is \gl\ invariant, the integral of
$\varphi_{\<e,n\>}$ is an intertwiner from $\<e,n\> \otimes
\<e,n\>$ to the trivial representation. This proves that the
expectation value we computed has the desired transformation
behavior.

\subsection{Bulk-boundary 2-point function}

Now we want to compute the full bulk-boundary 2-point function.
It is quite useful to determine the general form of this 2-point
function first before we enter the detailed calculations. Let us
suppose for a moment that our calculations were guaranteed to give
a \gl\ covariant answer. Then it is clear that the bulk-boundary
2-point function can be written as
\begin{eqnarray} \label{2pt}
\< \Psi_{\<2e',2n'\>}(0)\, \Phi_{\<-e,-n+1\>}(iy,-iy)\>
& = & \sum_{\nu=0,1} C_{\nu}(e)
  \frac{\<\psi_{\<2e',2n'\>} \, \varphi_{\<-e,-n+1\>}\>_\nu}
  {|y|^{2 \Delta_\nu} } \\[2mm]
  \text{ where }\quad   \Delta_0 \ = \
   \frac{2e}{k}\bigl(2n-1+\frac{e}{k}\bigr)
    & & \text{and}\qquad
    \Delta_1 \ = \ \frac{2e}{k}\bigl(2n-\frac{1}{2}+
   \frac{e}{k}\bigr)\ .
\end{eqnarray}
The structure constants $C_\nu(e)$ are not determined by the \gl\
symmetry. We will calculate them perturbatively below (see eqs.\
\eqref{C0} and \eqref{C1} below). The expressions in the numerator
on the right hand side are certain \gl\ intertwiners which are
defined by
\begin{eqnarray}  \label{int}
\<\psi_{\<2e',2n'\>} \, \varphi_{\<-e,-n+1\>}\>&\!\!\!=&
\hspace*{-3mm} \int d\mu\, \psi_{\<2e',2n'\>} \,
\phi_{\<-e,-n+1\>}
 =:\sum_{\nu=0,1} \<\psi_{\<2e',2n'\>} \,
  \varphi_{\<-e,-n+1\>}\>_\nu\quad   \\[2mm]
\text{where} \quad  \quad & & \hspace*{-5mm}
   \<\psi_{\<2e',2n'\>} \,
  \varphi_{\<-e,-n+1\>}\>_\nu \ = \
 \delta(e-e')\delta(n-n'-\nu/2)\  G_\nu
\label{intnu}
\end{eqnarray}
is the part of the full integral that contains the factor
$\delta(n-n'-\nu/2)$. Understanding the previous formulas requires
some input from the representation theory of \gl\ (see e.g.
\cite{Schomerus:2005bf} for all necessary details). Let us start
with the matrix $\varphi_{\<-e,-n+1\>}$. Under the  twisted
adjoint action of \gl\ this multiplet transforms in the tensor
product
$$\<-e,-n+1\> \otimes \<-e,-n+1\>  \ = \
  \< -2e,-2n + 2\> \oplus \< -2e,-2n+1\> \ \ . $$
Hence, there exist only two matrices $\psi_{\<2e',2n'\>}$ for
which the integral \eqref{int} does not vanish. These are the
matrices $\psi_{\<2e,2n\>}$ and $\psi_{\<2e,2n-1\>}$. The two
non-vanishing terms are used to define the the symbols
\eqref{intnu}. A similar analysis can now be repeated for the
fields in the WZNW model. We conclude immediately, that the
2-point function can only have two contributions. By \gl\
symmetry, these must be proportional to the intertwiners
\eqref{intnu}. The \gl\ symmetry, however, does not fix an overall
constant $C_\nu$ that can depend on the parameters of the fields.
Finally, the exponents $\Delta_\nu$ are simply determined by the
conformal dimensions of bulk and boundary fields. Let us point out
that the entire discussion leading to the expression \eqref{2pt}
is based on the global \gl\ symmetry. Since we have not yet shown
that our perturbative computations respect the action of \gl\ it
will be important to verify that the form of the 2-point function
comes out right.

In our perturbative computation, there are at most three fields
$c_\pm$ inserted and hence we only have to determine the expansion
terms for $s=0,1,2$. Contributions to the $\nu=0$ term in the
2-point function \eqref{2pt}, i.e. to the correlator with the
boundary field $\Psi_{\<2e,2n\>}$, can only come from $s=0$.
In fact, insertions of an interaction term - bulk or boundary -
would violate the conservation of $Y$-charge. Computation
without any insertion of an interaction are easily performed,
e.g.
\begin{equation}
\langle U^{11}_{\langle2e',2n'\rangle}(0)
  V^{00}_{\langle -e,-n+1\rangle}(iy,-iy)\rangle \ = \
   - \delta(n-n')\, \delta(e-e') |y|^{-4e/k(2n-1/2+e/k)}
\end{equation}
Here, we have introduced the notation $U^{\epsilon'\epsilon}$
and $V^{\epsilon'\epsilon}$ for matrix elements. The field
$U^{11}_{\<2e',2n'\>}$, for example, denotes the lower right
corner etc. The computation of the associated integral
\eqref{intnu} with $\nu=0$ is equally simple and allows us to
read off that
\begin{equation}  \label{C0}
C_0(e,n) \ = \ 1 \ \ .
\end{equation}
Let us note that there are other combinations of bulk and
boundary fields that can have a non-zero 2-point function
without any insertion of interactions. In all those cases
one may repeat the above calculation to find the same
coefficient $C_0 =1$, in agreement with \gl\ symmetry.

Next we would like to address the coefficient $C_1$ in the
expression \eqref{2pt}. $Y$-charge conservation implies that
its only contributions are associated with a single insertion
of the boundary interaction. This time, the computations are
slightly more involved. As an example we treat the
following 2-point function
\begin{equation}
    \begin{split}
             \langle U^{00}_{\langle2e',2n'\rangle}(0)
         V^{11}_{\langle-e,-n+1\rangle}&(iy,-iy)\, S^{\text{bdy}}_{\text{int}}   \rangle \, = \, \\[2mm]
         &= \, -\frac{\delta(n-n'-\frac12)\delta(e-e')}{|y|^{4\frac{e}{k}
   (2n-1+\frac{e}{k})}} \frac{y}{2\pi}\int du \ \frac{|u|^{2\alpha}}{|u^2+y^2|^{\alpha+1}} \\[2mm]
    &= \, -\frac{\delta(n-n'-\frac12)\delta(e-e')}{|y|^{4\frac{e}{k}
    (2n-1+\frac{e}{k})}} \frac{1}{2\pi}\int du \ |1+u^2|^{-\alpha-1} \\[2mm]
    &= \, -\frac{\delta(n-n'-\frac12)\delta(e-e')}{2|y|^{4\frac{e}{k}
    (2n-1+\frac{e}{k})}}2^{-2\alpha}
\frac{\Gamma(2\frac{e}{k}+1)}{\Gamma^2(\frac{e}{k}+1)} \\[2mm]
   &= \, -\frac{\delta(n-n'-\frac12)\delta(e-e')}{2|y|^{4\frac{e}{k}
  (2n-1+\frac{e}{k})}}
\frac{\Gamma(\frac{e}{k}+\frac12)}{\sqrt{\pi}\Gamma(\frac{e}{k}+1)}
\end{split}
\end{equation}
The second step is the substitution $u\rightarrow y/u$, then we can
apply \eqref{eq:1bdyint} which is a special case of the integral formula
in \cite{Fateev:2007}. The last step is the Euler doubling formula of
the Gamma function.
Comparison with the associated contribution to the minisuperspace
integral \eqref{int} gives
\begin{equation} \label{C1}
C_1(e) \ = \ \frac{\Gamma(e/k+1/2)}{\sqrt{\pi}\Gamma(e/k+1)}\ \ .
\end{equation}
Once more, one can perform similar computations with a single
insertion of a boundary interaction for other pairs of bulk
and boundary fields. All these calculations lead to the same
result for $C_1$, as predicted by \gl\ covariance.

At this point, we have computed all the data we were interested
in. But there are more contributions to the perturbative expansion
of the bulk-boundary 2-point function. As we stated above,
non-vanishing contributions arise from $s=0,s=1$ and $s=2$. We
have completely determined the $s=0$ term. At $s=1$, however, our
attention so far was restricted to the boundary interaction. The
other term with a single bulk insertion can also contribute since
it contains a product of only two $b_\pm$. Similarly, at $s=2$,
two insertions of the boundary interaction can lead to a
non-vanishing result. Products of bulk and boundary interactions
or two bulk interactions, on the other hand, involve too many
fields $b_\pm$ and vanish by simple zero mode counting. Hence, we
are left with two more terms to calculate, those arising from a
product of two boundary interactions $S^{\text{bdy}}_{\text{int}}$
and from a single bulk interaction $S^{\text{bulk}}_{\text{int}}$.
$Y$-charge conservation implies that the additional terms involve
a factor $\delta(n-n'-1)$. Such a term, if present,  would be
inconsistent with the global \gl\ symmetry. Our task therefore is
to show that the sum of the two aforementioned contributions
vanishes.

Let us begin with the computation of the term that arises from a
single insertion of the bulk interaction,
\begin{equation}\label{1bulk}
    \begin{split}
        \langle U^{11}_{\langle2e,2n-2\rangle}&(0)
    V^{11}_{\langle -e,-n+1\rangle}(iy,-iy)\ S^{\text{bulk}}_{\text{int}}
  \rangle \ \sim  \\[2mm]
        \sim \ & y^{-2\frac{e}{k}(4n-3+2\frac{e}{k})}\,
       \frac{y^3}{k\pi} \int_{UHP} d^2z\
      |z^2+y^2|^{-2(\frac{e}{k}+1)}|z^2|^{2\frac{e}{k}-1}(z-\bar{z}) \\[2mm]
%        = \ &  y^{-2\frac{e}{k}(4n-3+2\frac{e}{k})}\,
%    \frac{-i}{k\pi}\int_{LHP} d^2z\
%    |z^2+1|^{-2(\frac{e}{k}+1)}|z-\bar{z}| \\[2mm]
%        = \ &  -y^{-2\frac{e}{k}(4n-3+2\frac{e}{k})}\, \frac{\pi}
%      {k2^{4\frac{e}{k}}}
%       \frac{\Gamma(2\frac{e}{k}+1/2)\Gamma(2\frac{e}{k})}
%       {\Gamma(3/2)\Gamma^2(1+\frac{e}{k})
%        \Gamma^2(1/2+\frac{e}{k})}\\[2mm]
        = \ & -y^{-2\frac{e}{k}(4n-3+2\frac{e}{k})} \frac{1}{e\sqrt{\pi}}
    \frac{\Gamma(2e/k+1/2)}{\Gamma(2\frac{e}{k}+1)}\\[2mm]
    \end{split}
\end{equation}
We have been a bit sloppy here by setting the parameters the
parameters $2e'=2e$ and $2n'-2$ to the values at which the
expectation value has a non-vanishing contribution. Strictly
speaking, this quantity is divergent, but the divergence is an
overall (volume) factor $\delta(0)$ which we suppressed
consistently. In the first equality we simply inserted the
relevant free field correlator. After the substitution
$z\rightarrow y/z$, the integral over the insertion point $u$ of
the boundary interaction can be evaluated using an integral
formula from \cite{Fateev:2007} (see also \eqref{eq:1bulkint}). Finally, the
answer is simplified by means of Euler's doubling formula for
Gamma functions.

Next we turn to the contributions coming from two boundary interactions.
Since the corresponding free field correlator is slightly more involved
in this case, we state an expression for the fermionic contribution
before going into the actual computation,
\begin{equation}
    \begin{split}
        \langle\bp(u_1)C(u_1)&\bp(u_2)C(u_2)(\cp-\cm)(0)C(0)
    \cp(-iy)\cm(iy)\rangle^{\text {F}} \ = \\[2mm]
        = \ &\frac{-4\pi y^3(u_2-u_1)}{u_1u_2(u_1^2+y^2)(u_2^2+y^2)}
        \Bigl[\sign(u_2-u_1)-\sign(u_2)+\sign(u_1)\Bigr]\ .\\
    \end{split}
\end{equation}
This result is inserted to compute
\begin{equation}\label{2bdy}
    \begin{split}
        \langle  U^{11}_{\langle2e,2n-2\rangle}&(0)
  V^{11}_{\langle -e,-n+1\rangle}(iy,-iy)\
 \left(S^{\text{bdy}}_{\text{int}}\right)^2 \rangle
   \ \sim \\[2mm]
        = \ & y^{-2\frac{e}{k}(4n-3+2\frac{e}{k})}
        \frac{y^3}{\pi k}\int du_1du_2\
      |u_1^2+y^2|^{-e/k-1}|u_2^2+y^2|^{-\frac{e}{k}-1}
      |u_1^2|^{\frac{e}{k}-1}\\[2mm]
        \ & |u_2^2|^{\frac{e}{k}-1}(u_2-u_1)
   \Bigl[\sign(u_2-u_1)-\sign(u_2)+
      \sign(u_1)\Bigr] \\[2mm]
        = \ &  y^{-2\frac{e}{k}(4n-3+2\frac{e}{k})}
  \frac{1}{\pi k}\int dx_1dx_2\
    |x_1^2+1|^{-\frac{e}{k}-1}|x_2^2+1|^{-\frac{e}{k}-1}|x_1-x_2| \\[2mm]
%        = \ &  y^{-2\frac{e}{k}(4n-3+2\frac{e}{k})}
%       \frac{\pi}{k}2^{-4\frac{e}{k}+1}
%            \frac{\Gamma(2\frac{e}{k}+\frac12)\Gamma(2)
%    \Gamma(2\frac{e}{k})}
% {\Gamma(\frac{3}{2})\Gamma^2(1+\frac{e}{k})\Gamma^2(\frac12+
%     \frac{e}{k})}\\[2mm]
            = \ &  y^{-2\frac{e}{k}(4n-3+2\frac{e}{k})}
              \frac{2}{e\sqrt{\pi}}
  \frac{\Gamma(2\frac{e}{k}+\frac12)}{\Gamma(2\frac{e}{k}+1)}\\
    \end{split}
\end{equation}
The integral in the fourth line is again evaluated with a special
case of the integral formula of Fateev and Ribault
\eqref{eq:2bdyint}. Putting the results of eqs.\ \eqref{1bulk} and
\eqref{2bdy} together we arrive at
\begin{equation}
  \langle U^{11}_{\langle2e',2n'\rangle}(0)
   V^{11}_{\langle -e,-n+1\rangle}(iy,-iy) \left( S^{\text{bulk}}_{\text{int}}
    + \frac{1}{2!} \left( S^{\text{bdy}}_{\text{int}}\right)^2 \right)
     \  \rangle \ = \ 0 \ \ ,
\end{equation}
in agreement with \gl\ covariance of the 2-point function. Thereby,
we have now established the formula \eqref{2pt} through our
perturbative computations.

Before we leave the subject of bulk boundary 2-point functions, we
would like to make a few comments on the cases when $e/k$ is an
integer multiple of $1/2$. Consider inserting a bulk
vertex operator with $e$ momentum $e=-mk-k/2-k\varepsilon$ and
sending $\varepsilon$ to zero. In the limit, the second term of
eq.\ \eqref{2pt} develops a logarithmic singularity,
\begin{equation}
    \begin{split} \label{2ptlog}
        C_1(-mk-k/2-k\epsilon)|y|^{-\Delta_1} \ &= \
        \frac{(-1)^m}{m!\Gamma(-m+1/2) |y|^{2\Delta}} (\mathcal{Z} +
        \tilde{\Delta}\ln|y| + o(\epsilon)) \\[2mm]
        \text{where}\qquad \mathcal{Z} \ &= \
        \frac{1}{\epsilon}+\Psi(-m)-\Psi(-m+1/2) \ , \\[2mm]
        \Delta \ &= \ -(2m+1)(2n-m-1) \ . \\[1mm]
    \end{split}
\end{equation}
and $\tilde{\Delta} = 4n-4m-3$.  Here, $\Psi$ is the usual
Di-gamma function. The form of our bulk-boundary 2-point function
\eqref{2ptlog} resembles a similar expression in
\cite{Gaberdiel:2006pp}. A link between boundary
correlation functions of symplectic fermions and the
corresponding correlators in the \GL\ WZNW model may be
established following ideas in \cite{LeClair:2007aj}.

\subsection{Boundary 3-point functions}

The second object of interest for us is the boundary 3-point
function. Before we get there, we have to turn our attention to an
important detail that we glossed over in the previous subsection.
We recall that our $2\times 2$ matrices $\Psi_{\<e,n\>}, e \neq k
\mathbb{Z},$  of boundary fields contain two irreducible
multiplets $\< e,n\>$ under the unbroken global \gl\ symmetry.
These two multiplets have opposite fermion number, i.e.\ the state
with lower eigenvalue of $N$ is bosonic for one of them and
fermionic for the other. In general, the two multiplets are
allowed to have different couplings to the other fields in the
theory. When we studied bulk-boundary 2-point function, only one
of the two multiplets from each of the $2\times2$ matrices
$\Psi_{\<2e,2n\>}$ and $\Psi_{\<2e,2n-1\>}$ could have a
non-vanishing overlap with the bulk field $\Phi_{\<-e,-n+1\>}$,
simply because of fermion number conservation. Hence, the
bulk-boundary 2-point functions were parametrized by two
non-vanishing structure constants $C_\nu(e)$ rather than four. For
boundary 3-point functions, however, the distinction becomes
important. Consequently, we introduce the symbols
\begin{equation}
\begin{split} U^0_{\<-2e,-2n+1\>}(u)\ & =
  \ e^{ieX+inY}\, (\  1\  , \ \cp-\cm\ )  \\[2mm]
              U^1_{\<-2e,-2n+1\>}(u)\ & =
\ e^{ieX+inY}\, (\, C , (\cp-\cm)\, C\, )
\end{split}
\end{equation}
for the first and second row of the matrix \eqref{vertU}. The same
notation is used for the rows of the matrices $\psi$ of functions
and $\Psi$ of boundary fields.

Let us now begin with the 3-point function of three fields from
the first multiplet $\Psi^0$. These acquire contributions
exclusively from a single insertion of the boundary interaction. A
non-vanishing correlator requires that the parameters $e_i$ of the
three fields sum up to $\tilde e = e_1+e_2 + e_3 = 0$ and
similarly that $\tilde n = n_1 + n_2 + n_3 = 1$. Using the
integral formulas from Appendix A, the 3-point function of
fields $\Psi^0$ in the regime $0 < x < 1$ is found to be
\begin{equation}\label{3t000}
    \begin{split}
     &\langle \Psi^{0\epsilon_1}_{\langle -2e_1,-2n_1+1\rangle}(0)
   \Psi^{0\epsilon_2} _{\langle -2e_2,-2n_2+1\rangle}(1)
   \Psi^{0\epsilon_2}_{\langle -2e_3,-2n_3+1\rangle}(x)\rangle \ =
   \delta(\tilde e)\, \delta (\tilde n -1) \, \delta(\tilde \epsilon-2)\, \times \\[4mm]
        &\qquad\qquad  \times\  x^{2\Delta_{13}}(1-x)^{2\Delta_{23}}\
        \frac{\pi}{i} \frac{s(\alpha_1)+s(\alpha_2)+s(\alpha_3)}
   {s(\alpha_1)s(\alpha_2)s(\alpha_3)
 \Gamma(\alpha_1+\epsilon_1)\Gamma(\alpha_2+\epsilon_2)
 \Gamma(\alpha_3+\epsilon_3)}\\
    \end{split}
\end{equation}
where we defined the parameters $\alpha_i$ by $\alpha_i =2e_i/k$
and introduced the short-hands $s(z)$ and $\tilde \epsilon$ for
$s(z) = \sin(\pi z)$ and $\tilde \epsilon  = \sum \epsilon_i$. The
conformal weights are given by
$$ \Delta_{ij}\ = \ (n_i-1/2)\alpha_j+(n_j-1/2)
 \alpha_i+\alpha_i\alpha_j\ . $$
In the limit $k \rightarrow \infty$ the function $s(\alpha_i)$ can
be approximated by $s(\alpha) \sim 2 \pi e_i/k $ and the entire
3-point function is seen to vanish due to the conservation of $e$
momentum. This is consistent with the minisuperspace theory. In
fact, the corresponding  integral of functions on our brane is
easily seen to vanish,
$$ \langle \psi^{0\epsilon_1}_{\langle -2e_1,-2n_1+1\rangle}
   \psi^{0\epsilon_2} _{\langle -2e_2,-2n_2+1\rangle}
   \psi^{0\epsilon_2}_{\langle -2e_3,-2n_3+1\rangle}\rangle
    \ = \ 0\ .
    $$
This is so
because integration with the Haar measure needs a product of two
different fermionic zero modes in order to give a non-zero result.
Our functions $\psi^0$, however, only contain the zero mode
$\eta_+ -\eta_-$.

Let us now move on to discuss the 3-point in the case where a
single field from the second multiplet $\Psi^1$ is inserted.
Contributions to such correlators arise only from the leading term
$s=0$ of the perturbation series (see below). The result is
therefore straightforward to write down
\begin{equation}\label{3pt001}
    \begin{split}
    & \langle \Psi^{0\epsilon_1}_{\langle -2e_1,-2n_1+1\rangle}(0)
    \Psi^{0\epsilon_2}_{\langle -2e_2,-2n_2+1\rangle}(1)
    \Psi^{1\epsilon_3}_{\langle -2e_3,-2n_3+1\rangle}(x)\rangle
    \ = \ \qquad\qquad \qquad \\[3mm] & \qquad \qquad \qquad
     \qquad \qquad \qquad  \ = \ \delta(\tilde e)\,
       \delta (\tilde n - 1/2) \,
         \delta(\tilde \epsilon-1) \
        x^{2\Delta_{13}}(1-x)^{2\Delta_{23}}\ . \\
    \end{split}
\end{equation}
This coupling in independent of the level $k$ and it matches the
minisuperspace answer which is non-zero because the multiplet
$\psi^1$ contains both fermionic zero modes.

The most interesting 3-point coupling appears when we insert two
fields from the second multiplet $\Psi^1$. Once more,
non-vanishing terms can only arise from the insertion of a single
boundary interaction. They can be worked out with the help of
integral formulas in Appendix A,
\begin{equation}\label{3pt011}
    \begin{split}
        &\langle \Psi^{0\epsilon_1}_{\langle -2e_1,-2n_1+1\rangle}(0)
        \Psi^{1\epsilon_2}_{\langle -2e_2,-2n_2+1\rangle}(1)
        \Psi^{1\epsilon_3}_{\langle -2e_3,-2n_3+1\rangle}(x)\rangle
         \ = \ \delta(\tilde e)\, \delta (\tilde n - 1) \,
         \delta(\tilde \epsilon-2)\ \times \\[4mm]
        &\qquad  \times \ \frac{2\pi^2i}{k}\,
        x^{2\Delta_{13}}(1-x)^{2\Delta_{23}}\
       \frac{s(\alpha_1)-s(\alpha_2)-s(\alpha_3)}
{s(\alpha_1)s(\alpha_2)s(\alpha_3)
\Gamma(\alpha_1+\epsilon_1)\Gamma(\alpha_2+\epsilon_2)
 \Gamma(\alpha_3+\epsilon_3)}\ \ . \\
    \end{split}
\end{equation}
Note that the factor $\sim 1/k$ in the first term of the second
row is necessary in order for the whole expression to scale to a
finite value as we send the level $k$ to infinity. The expression
that arises in this limit can be checked easily in the
minisuperspace theory.

There remains one more case to consider, namely the 3-point
function for three fields from the second multiplet $\Psi^1$.  It
is given by
\begin{equation}\label{3pt111}
    \begin{split}
    & \langle \Psi^{1\epsilon_1}_{\langle -2e_1,-2n_1+1\rangle}(0)
    \Psi^{1\epsilon_2}_{\langle -2e_2,-2n_2+1\rangle}(1)
    \Psi^{1\epsilon_3}_{\langle -2e_3,-2n_3+1\rangle}(x)\rangle
    \ = \\[3mm]
     & \qquad \qquad\qquad \ = \ \delta(\tilde e)\, \delta (\tilde n - 1/2) \,
         \delta(\tilde \epsilon-1) \frac{2\pi}{k}\
        x^{2\Delta_{13}}(1-x)^{2\Delta_{23}}\ \ . \\
    \end{split}
\end{equation}
As in the previous formula  \eqref{3pt011}, the result contains a
factor $1/k$. Consequently, the 3-point coupling on the right hand
side of  eq.\ \eqref{3pt111} vanishes at $k \sim \infty$, in
agreement with the associated minisuperspace computation.

The last result \eqref{3pt111} was obtained without any insertion
of bulk or boundary interactions, though naively one might expect
to see contributions from one bulk or two boundary insertions. A
similar comment applies to the second case \eqref{3pt001} above.
It is indeed true that the insertion of $S^{\text
bulk}_{\text{int}}$ or $(S^{\text{bdy}}_{\text{int}})^2$ both lead
to non-vanishing expressions. But, as in the case of the bulk
boundary 2-point functions, their sum vanishes, i.e.\
$$\langle\,  U^{\epsilon'_1\epsilon_1}_{\langle e_1,n_1\rangle}(0)\,
 U^{\epsilon'_2\epsilon_2}_{\langle e_2,n_2\rangle}(1)\,
 U^{\epsilon'_3\epsilon'_3}_{\langle e_3,n_3\rangle}(u)\,
 \left( S^{\text{bulk}}_{\text{int}} + \frac{1}{2!} \left(
    S^{\text{bdy}}_{\text{int}}\right)^2\right)
    \rangle\ = \ 0 \ .
$$
The result is trivially fulfilled for $\tilde \epsilon' = 0,2$. It
requires rather elaborate computations when $\tilde \epsilon' =
1,3$. These can be performed with the help of the integral
formulas (\ref{eq:A3}-\ref{eq:A5}) we list in Appendix A.

Before closing this section we would like to add two more
comments. The first one concerns the logarithmic singularities
that appear in the 3-point functions whenever one of the
parameters $2e_i$ is an integer multiple of $k$. If we
consider joining two open strings with $e$ momentum $e_1 = e -
\varepsilon/2$ and $e_2= - e - \varepsilon/2$, for example, and
send $\varepsilon$ to zero, we obtain
\begin{equation}
    \begin{split} \label{eq:atypical}
        &\langle \Psi^{00}_{\langle -2e+\varepsilon,-2n_1+1\rangle}(0)
         \Psi^{11}_{\langle 2e+\varepsilon,-2n_2+1\rangle}(1)
         \Psi^{11}_{\langle -2\varepsilon,-2n_3+1\rangle}(u)\rangle \ \sim
   \\[2mm]
        &\qquad  \ \sim \ u^{2\Delta}(1-u)^{-2\Delta}
   \,  \delta (\tilde n-1) \
    \bigl(\mathcal{Z}+\mathcal{R}(\alpha)+A_{23}\ln|1-u|+A_{13}\ln|u|+ o(\varepsilon)\bigr)\\[4mm]
    &\qquad\text{where} \qquad \mathcal{Z} \ = \ \frac{1}{\varepsilon}+ \frac{4\varepsilon\gamma}{k}
     \qquad , \qquad \mathcal{R}(\alpha)\ = \ -2\pi\frac{1+c(\alpha)}{ks(\alpha)}  \\[2mm]
    & \qquad  A_{13} \ = \ \frac{1}{k}(2n_1-n_3-1/2+2\alpha)\qquad , \qquad
     A_{23}\ = \ \frac{1}{k}(2n_2-n_3-1/2-2\alpha)\\[2mm]
    \end{split}
\end{equation}
and $\Delta \ = \ \alpha(n_3-1/2)$. The function $c(\alpha)$
stands for $c(\alpha) = \cos( \pi \alpha)$ and $\gamma$ is the
Euler-Mascheroni constant. In the limit $\varepsilon \rightarrow 0$, the
constant $\mathcal{Z}$ diverges. This divergency can be regularized by
adding to $\Psi^{11}$ an appropriate field from the socle of the
involved atypical multiplet. In the following, we shall assume that
$\mathcal{Z}$ has been set to zero.

Our final comment deals with an interesting quantum symmetry of
the boundary 3-point functions. As in the bulk sector
\cite{Schomerus:2005bf}, the boundary 3-point function is periodic
under shifts of the $e$-momentum, in the following sense,
\begin{equation} \nonumber
    \begin{split}
          &\langle \Psi^{\epsilon_1\epsilon_1'}_{\langle 
                 -2e_1,-2n_1+1\rangle}(0)
          \Psi^{\epsilon_2\epsilon_2'}_{\langle 
                 -2e_2,-2n_2+1\rangle}(1)
          \Psi^{\epsilon_3\epsilon_3'}_{\langle 
                 -2e_3,-2n_3+1\rangle}(x)\rangle\ = \\[4mm]
          &\qquad(1-u)^{2n_3-1}u^{1-2n_3}
          \langle \Psi^{\epsilon_2\epsilon_2'}_{\langle 
                 -2e_1+k,-2n_1\rangle}(1)
          \Psi^{\epsilon_1\epsilon_1'}_{\langle 
                 -2e_2-k,-2n_2+2\rangle}(0)
          \Psi^{\epsilon_3\epsilon_3'}_{\langle 
                  -2e_3,-2n_3+1\rangle}(x)\rangle\ \ . \\
      \end{split}
\end{equation}
Further shifts by multiples of $\pm k$ can also be considered,
but necessarily involve inserting descendants of the tachyon
vertex operators. Our observation proves that the boundary \GL\
model for volume filling branes possesses spectral flow symmetry.
Shifts by integer multiples of the level $k$ are a symmetry of 
the affine representation theory. In principle, this symmetry 
could be broken by the boundary structure constants. The previous
formula asserts that, like in the bulk sector, the boundary 
operator product expansions preserve the spectral flow symmetry. 
The same is true for the bulk-boundary operator product 
expansions.

\section{Correlation functions involving atypical fields}

Throughout the last few sections we have learned how to compute
correlation functions of bulk and boundary tachyon vertex
operators for a volume filling brane in the \GL\ WZNW model. We
now want to add a few comments on a particular set of correlation
functions that are essentially not effected by the interaction and
hence can be derived without cumbersome calculations. These will
include a non-vanishing annulus amplitude. We shall use the latter
to perform a highly non-trivial test on the proposed boundary
state of volume filling branes \cite{Creutzig:2007jy}.

\subsection{Correlators for special atypical fields}

In the previous sections we developed a first order formalism for
computations of correlation functions in the \GL\ WZNW model. Very
special correlators, however, can also be computed in the original
formulation. To begin with, let us explain the main idea at the
example of bulk correlators. We recall that the bulk action of the
\GL\ model is given by
\begin{equation}\label{eq:bulkSWZW}
    \begin{split}
        S_{\text{bulk}}\ =\ &-\frac{k}{4\pi i}\int_\Sigma d^2z\
        \left(
        \del X\bar{\del}Y+\del
        Y\bar{\del}X+2e^{iY}\del\cp\bar{\del}\cm\right)
    \end{split}
\end{equation}
The path integral is evaluated with the \gl\ invariant
measure \eqref{WZNWmeasure} on the space of fields.
A glance at the interaction term of the WZNW model and the measure
suggests to introduce the new coordinates $\chi_\pm=e^{iY/2}\cpm$.
After this substitution,  the path integral measure is the canonical
one,
\begin{equation}
    d\mu_{\text{WZW}} \ \sim \ \mathcal{D}X\mathcal{D}Y\mathcal{D}
     \chim\mathcal{D}\chip\ .
\end{equation}
Our bulk action $S_{\text{bulk}}= S_0 + Q$, on the other hand,
splits naturally into a free field theory $S_0$ and an interaction
term $Q$ where
\begin{equation}\label{eq:bulk}
    \begin{split}
    \qquad S_0\ =\ &-\frac{k}{4\pi i}\int_\Sigma d^2z\ \left( \del X\bar{\del}Y+
            \del Y\bar{\del}X+2\del\chip\bar{\del}\chim\right) \\[2mm]
    \qquad  \ \ Q\ = \ &\frac{k}{4\pi i}\int_\Sigma d^2z\ \left(i\chip\bar{\del}\chim\del Y+
     i\del\chip\chim\bar{\del}Y+\chip\chim\del Y\bar{\del}Y\right)  \ .\\
    \end{split}
\end{equation}
Due to the complicated form of $Q$, treating the WZNW model as a
perturbation by the interaction terms in $Q$ is not too useful for
most practical computations. Under very special circumstances,
however, the split into $S_0$ and $Q$ allows for a very
interesting conclusion. Observe that each term in the interaction
$Q$ contains at least one derivative $\partial Y$ or $\bar
\partial Y$. In our free field theory $S_0$, the only
non-vanishing contractions involving derivatives of $Y$ are those
with the field $X$. Hence, we can simply ignore the presence of
$Q$ for all correlation functions of tachyon vertex operators that
do not involve $X$. In other words, correlation functions of
fields without any $X$-dependence are given by their free field
theory expressions! This had already been observed in the results
of \cite{Schomerus:2005bf}. Our split of the action in $S_0$ and
$Q$ provides a rather simple and general explanation. Let us
stress again that this split is not helpful for any other
computation involving more generic typical fields.

It is clear that all this is not restricted to the bulk theory. In
fact, we can use the same substitution for the boundary terms of
the action \eqref{eq:SWZW},
\begin{equation}\label{eq:bdytwisted}
    \begin{split}
        S_{\partial 0}\ =\ &\frac{k}{8\pi i}\int_\Sigma du\ (\chip+\chim)
          \del_u(\chip+\chim)\ . \\
    \end{split}
\end{equation}
Since $S_{\partial 0}$ is quadratic in the fields $\chi_\pm$, it
gets added to the free bulk action $S_0$, i.e.\ we now work with a
free field theory on the upper half plane whose action is given by
$S_0+S_{\partial 0}$. There is no additional boundary contribution
to the bulk interaction $Q$. In the free theory, the fields
$\chi_\pm$ satisfy Neumann gluing conditions of the following
simple form,
\begin{equation}\label{eq:bdyfermions}
    \del \chipm(z,\bar{z}) \ = \ \mp \bar{\del} \chi_{\mp}(z,\bar{z})
    \qquad \text{for}\qquad z \ = \ \bar{z} \ .
\end{equation}
The gluing condition implies that fermions of the free boundary
theory are contracted as follows,
\begin{equation}
    \begin{split}
        \chim(z,\bar{z})\chip(w,\bar{w}) \ &\sim \ \frac{1}{k}\ln |z-w|^2 \ ,
        \\[2mm]
        \chipm(z,\bar{z})\chipm(w,\bar{w}) \ &\sim \
         \frac{1}{k}\ln (\bar{z}-w)-\frac{1}{k}\ln (\bar{w}-z) \ . \\
    \end{split}
\end{equation}
The bosonic fields $X,Y$ also obey simple Neumann boundary
conditions so that the evaluation of correlators in the free field
theory $S_0 + S_{\partial 0}$ is straightforward. Taking the
interaction $Q$ into account is a difficult task unless none of
the vertex operators in the correlation function contain the field
$X$. If all field are $X$ independent, then the correlator is
simply given by the free field theory formula, just as in the bulk
theory above.

One may apply the observation in the previous paragraph to the
evaluation of boundary 3-point functions of three atypical fields
for the volume filling brane. Note that we did not spell out a
formula for this particular correlator before. In principle, it
can be computed in the first order formalism, but the
corresponding calculation requires some care. Our new approach
allows to write down the result right away.
% VS: lets spell it out
We shall discuss another interesting application of our new
approach to atypical correlation functions in the next subsection.
Let us mention in passing that we expect similar results to hold
for the completely atypical sectors in all $GL(N|N)$ and
$PSL(N|N)$ WZNW models. This will be discussed in more detail
elsewhere.

\subsection{Twisted boundary state and modular bootstrap}

In our previous paper \cite{Creutzig:2007jy}, we proposed a
formula for a boundary state of volume filling brane on \GL. The
usual annulus amplitude for this boundary state was trivially
zero, in agreement with the observation that open string states
are perfectly paired. In fact, as we have mentioned at various
places throughout this note, for each multiplet $\< e,n\>$ of
boundary fields there exists one with opposite parity.
Contributions of such pairs to the boundary partition function
cancel each other, leading to a vanishing boundary partition
function.

In order to construct a non-trivial quantity on the annulus, we
need to insert some fermionic zero modes, see e.g.\
\cite{Creutzig:2006} for similar tests in the simpler $bc$ ghost
system. Previously, we have not been able to compute such
quantities in the \GL\ WZNW model. We can now fill this gap! Let
us anticipate that only atypical bulk fields couple to the volume
filling brane. Hence, if we insert fermionic zero modes through
some atypical bulk field, the entire amplitude is built from
atypical terms and should be computable through a simple free
field formalism, as explained in the previous subsection. Let us
see now how the details of this calculation work out.

To begin with, let us review the construction of the boundary
state $|\Omega\>$ for the volume filling brane. With the help of
our free field realization, the formula becomes quite explicit. We
shall start from the boundary state $|\Omega\>_0$ of the free
theory. This state clearly factorizes into a product of a bosonic
$|\Omega,B\rangle_0$ and a fermionic $|\Omega,F \rangle_0$
contribution. The latter two obey the following gluing conditions
\begin{equation} \label{glue01}
    (X_n + \bar X_{-n})\, |\, \Omega,B\rangle_0\ =\
    (Y_n + \bar Y_{-n})\, |\, \Omega,B\rangle_0 \ =\ 0
\end{equation}
and
\begin{equation}\label{glue02}
    ( \chi^\pm_n \mp \bar \chi^{\mp}_{-n})
    \, |\, \Omega,F\rangle_0\ =\  0\ \ .
\end{equation}
Here, $X_n$ and $\bar X_n$ are the modes of the currents $i
\sqrt{k} \del X$ and $i\sqrt{k}\bar \del X$ etc. Up to
normalization, there exists a unique solution for these linear
constraints. For the bosonic and the fermionic sector, they are
given by the following coherent states,
\begin{eqnarray} \label{bound01}
|\, \Omega,B\rangle_0 & = & \exp \left( - \sum_{n=1}^\infty
\frac{1}{n} (Y_{-n} \bar X_{-n} + X_{-n} \bar Y_{-n}\right)
\,|0,0\rangle_B \\[2mm] \label{bound02}
|\, \Omega,F\rangle_0 & = & \exp \left( - \sum_{n=1}^\infty
\frac{1}{n} (\chi^+_{-n} \bar \chi^+_{-n} - \chi^-_{-n} \bar
\chi^-_{-n}\right)|0,0\rangle_F\ \ .
\end{eqnarray}
Here, $|0,0\rangle$ denote the vacua in the bosonic and the
fermionic theory. The product of the two components is the
boundary state of the free field theory, before the interaction is
taken into account. We now include the effects of the interaction
by multiplying the free boundary state with the exponential of the
interaction $Q$,
\begin{equation} \label{BS}
    |\, \Omega\rangle \ = \ \mathcal{N} e^{Q}\,  |\, \Omega\rangle_0 \ = \
     \mathcal{N}\ \left( \sum_{n=0}^\infty\
    \frac{Q^n}{n!}\right)  \ |\, \Omega,B\rangle_0 \times\,  |\, \Omega,F\rangle_0 \ ,
\end{equation}
where $\mathcal{N}=\sqrt{\pi/2 i}$ is a normalization constant.
The operator $Q$ is defined as in eq.\
\eqref{eq:bulk}, but with the integration restricted to the
interior of the unit disc. It is possible to check that $\exp{Q}$
rotates the gluing conditions from the free field theory relations
\eqref{glue01} and \eqref{glue02} to their interacting
counterparts (see \eqref{eq:bdyeom}). The dual boundary state is
constructed analogously.

Our main aim now is to compute some non-vanishing overlap of the
twisted boundary state $|\Omega\rangle$. This requires the
insertion of the invariant bulk field $\Phi^{11}_{\<0,0\>}  =
\chim\chip $, i.e.\ we are going to study
\begin{equation}
    \begin{split}\label{eq:amplitude}
 Z_\Omega(q,z) \ := \ \langle\Omega\, |\, \tilde q^{L_0^c}(-1)^{F^c}\,
   \tilde z^{N_0^c}\, \Phi^{11}_{\<0,0\>}
         \, |\, \Omega\rangle\ ,
    \end{split}
\end{equation}
where $L_0^c = (L_0+\bar{L}_0)/2$ and $N_0^c = (N_0-\bar{N}_0)/2$
are obtained from the zero modes of the Virasoro field and the
current $N$. The corresponding expressions are standard, see e.g.
\cite{Schomerus:2005bf}. Our parameters $\tilde q$ and $\tilde z$
are defined in terms of $\mu,\tau$ through $\tilde q = \exp (-2\pi
i/\tau)$ and $\tilde z = \exp(2\pi i \mu/\tau)$.  We are now going
to argue that the computation of $Z_\Omega$ can be reduced to a
simple calculation in free field theory, i.e.\
\begin{equation}
    \begin{split}\label{eq:amplitudetilde}
        \langle\Omega\, |\, \tilde q^{L_0^c}(-1)^{F^c}\, \tilde z^{N_0^c}
       \,  \Phi^{11}_{\<0,0\>} \, |\, \Omega\rangle\ = \ \mathcal{N}^2\
    _0\langle\Omega\, |\, \tilde q^{{L}_0^c}(-1)^{F^c}\, \tilde
    z^{{N}_0^c}\, \Phi^{11}_{\<0,0\>} \, |\, \Omega\rangle_0\ .
    \end{split}
\end{equation}
The reasoning goes as follows. In a first step we write the
interacting boundary state as a product of the interaction term
$\exp Q$ and the free boundary state $|\Omega\>_0$. Next we
observe that all bosonic operators in between the two boundary
states involve derivatives  such as  $\del X$ etc. Hence, we can
use the gluing conditions \eqref{glue01} to express all these
terms through $Y_n$ and $X_n$. The modes $\bar Y_n$ and $\bar X_n$
of the anti-holomorphic derivatives only appear in the
construction \eqref{bound01} of the free bosonic boundary state
$|\Omega,B\>_0$. A non-vanishing term requires that the number of
$\bar X_n$ equals the number of $\bar Y_{-n}$. But since the $\bar
X_{-n}$ and $\bar Y_{-n}$ come paired with their holomorphic
partners $Y_{-n}$ and $X_{-n}$ in the boundary state, the operator
in between $_0 \< \Omega|$ and $|\Omega\>_0$ must have equal
numbers for $X_n$ and $Y_n$ modes in order for the corresponding
term not to vanish. In $Q$, all terms have an excess of $Y$ modes.
Since no term in $L_0^c$ or $N_0^c$ can compensate this through an
excess of $X$-modes, we can safely replace $\exp Q$ by its zeroth
order term, i.e.\ $\exp Q \sim 1$.

The computation of the overlap \eqref{eq:amplitudetilde} in free
field theory is straightforward. In a first step, the amplitude is
split into a product of bosonic and fermionic terms. The bosonic
contribution is the same as for extended branes in flat
2-dimensional space. The fermionic factor involves an insertion.
Its evaluation is reminiscent of a similar calculation in
\cite{Creutzig:2006}. We can express the result through a single
character of the affine \gl\ algebra,
\begin{equation}
 Z_\Omega(q,z) \  = \ \mathcal{N}^2\ \
  \hat{\chi}_{{\cal P}_0}(-1/\tau,\mu/\tau)\
  =\ \frac{\pi}{k}\int de dn \ \frac{\hat{\chi}_{\langle e,n\rangle}
        (\tau,\mu)} {\sin(\pi e/k)} \ .
\end{equation}
The affine characters $\hat{\chi}$ along with their behavior under
modular transformations can be found in the Appendix A of
\cite{Creutzig:2007jy}. In order to achieve proper normalization
(see below) we have set ${\cal N}^2=\pi/2i$. Since the spectrum of
boundary operators on the volume filling brane is continuous, the
result involves some open string spectral density function. From
the result, this is read off as
\begin{equation} \label{SD}
 \rho(e,n) \ = \
\rho(e) \ = \ \frac{\pi}{k\sin(\pi e/k)}\ \ .
\end{equation}
We would expect $\rho$ to be encoded in the boundary 3-point
function of $\Psi_{\langle e,n\rangle}$, $\Psi_{\langle
-e,-n\rangle}$ with the special boundary field
$\Psi^{11}_{\<0,0\>}$. One possible 3-point function that
contains the required information is a particular case
of our more general formula \eqref{eq:atypical}, i.e.
\begin{equation}
    \begin{split}
    &\langle \Psi^{00}_{\langle e,n\rangle}(0) \Psi^{11}_{\langle -e,-n\rangle}
      (1)\Psi^{11}_{\<0,0\>} \rangle \ \sim \ \\[3mm]
        &\qquad\qquad\qquad\ \sim \ u^{2\Delta}(1-u)^{-2\Delta}
    \bigl(\mathcal{Z} +\mathcal{R}(-\pi e/k)+A_{23}\ln|1-u|+
    A_{13}\ln|u|\bigr)\ \ . \\[2mm]
    \end{split}
\end{equation}
All quantities that appear on the right hand side were introduced
in equation \eqref{eq:atypical}. The additive constant ${\cal Z}$
is not universal. It is naively infinite, but can be made finite
by a proper regularization prescription. We use the universal term
${\cal R}$ to determine the spectral density
\begin{equation}
    \frac{d}{de}\ln \mathcal{R}(-\pi e/k) \ = \
    \frac{\pi}{k} \frac{d}{d\alpha}\ln \frac{1+c(\alpha)}{s(-\alpha)}
     \ = \  \frac{\pi}{k\sin(\pi e/k)} \ = \  \rho(e) \ .
\end{equation}
Here, we have used that $\alpha =  e/k$, as before. The result
agrees with the expression \eqref{SD} that was obtained through
modular transformation of the overlap \eqref{eq:amplitudetilde}.
Thereby, we have now been able to subject our formula \eqref{BS}
for the boundary state of the volume filling brane to a strong
consistency check.

There is another somewhat weaker but still non-trivial test for
the boundary state that arises from the minisuperspace limit of
the boundary WZNW model. In fact, in the particle limit we find
that
\begin{equation}
    \tr(z^{\ad^\Omega_N}(-1)^F \psi^{11}_{\langle 0, 0\rangle})
    \ = \ \int dedn \ \frac{\chi_{\langle e,n\rangle}(z)}{e} \ = \
    \lim_{k\rightarrow\infty} Z_\Omega(q,z)\ .
\end{equation}
In the first step we simply evaluated the trace directly in the
minisuperspace theory. We then observed in the second equality
that the result coincides with the modular transform of the
overlap \eqref{eq:amplitudetilde} in the appropriate limit $k
\rightarrow \infty$.

\section{Conclusions and open problems}

In this note we have solved the boundary theory for the volume
filling brane on \GL. We achieved this with the help of a
Kac-Wakimoto-like representation of the boundary theory. The first
order formalism we developed in section 2 is similar to the one
used in \cite{Fateev:2007} for $AdS_2$ branes in the Euclidean
$AdS_3$. The main difference is that we were forced to introduce
an additional fermion on the boundary. Such auxiliary boundary
fermions are quite common in fermionic theories (see e.g.\
\cite{Warner:1995ay,Hosomichi:2004ph} and references therein).
With the help of our first order formalism we were then able to
set up a perturbative calculational scheme for correlation
functions of bulk and boundary fields. The main features of this
scheme are similar to the pure bulk case \cite{Schomerus:2005bf}.
In particular, for any given correlator, only a finite number of
terms from the expansion can contribute. We computed the exact
bulk-boundary 2-point functions and the boundary 3-point
functions, thereby solving the boundary conformal field theory of
volume filling branes on \GL\ explicitly. Finally, we proposed a
second approach to correlation functions of atypical fields. It
singles out a particular subsector of the bulk and boundary \GL\
WZNW model that is not affected at all by the interaction. Hence,
within this subsector, all quantities agree with their free field
theory counterparts. The insight was then put to use for a
calculation of a particular non-vanishing annulus amplitude in
section 5.2. Together with our previous results on boundary
3-point functions, we obtained a strong test for the boundary
state of the volume filling brane in the \GL\ WZNW model.

There are several obvious extensions that should be worked out. To
begin with, it would be interesting to set up an equally efficient
framework to calculate correlation functions for the boundary
theories of point-like localized branes. Unfortunately, we have
not succeeded to calculate correlators from a finite number of
contributions, as in the case of the volume filling brane. It is
possible to develop a Kac-Wakimoto-like presentation for
point-like branes using the boundary conditions of
\cite{Creutzig:2006} for the $bc$ system. But since the gluing
conditions of \cite{Creutzig:2006} identify derivatives of $c$
with $\bar b$ etc., zero mode counting does not furnish simple
vanishing results. Therefore, an infinite number of terms can
contribute to any given correlation function. On the other hand,
the second approach of section 5 does generalize to
point-like branes. Since the boundary spectrum on a single
point-like brane is purely atypical, some interesting quantities
can be computed. This applies in particular to the boundary
3-point functions on a single point-like brane.
Correlation functions involving boundary condition changing fields
or typical bulk fields, however, are not accessible along these
lines.

It is certainly interesting to investigate how much of our program
extends to higher supergroups. Encouraged by the recent
developments on the bulk sector \cite{Quella:2007hr}, it seems
likely that most of our constructions may be generalized, at least
to supergroups of type I. This includes the superconformal
algebras psl(N$|$N) and many other interesting Lie superalgebras
(see e.g. \cite{Frappat:1996pb} for a complete list). We believe
that in all these cases there exists one class of branes which can
be solved through some appropriate square root of the bulk
formalism. Taking the proper square root will certainly involve a
larger number of fermionic boundary fields. Our second approach to
atypical correlation functions may also be extended to higher
supergroups and it provides interesting insights on the atypical
subsector of the WZNW models. We plan to return to these issues in
a forthcoming publication.
\bigskip
\bigskip 
\bigskip

\noindent {\bf Acknowledgements:} We wish to thank Yasuaki Hikida,
Vladimir Mitev, David Ridout, Peter R\o nne and in particular 
Thomas Quella and Sylvain Ribault for interesting discussions and 
comments on issues related to this work and on the manuscript. 
This work was supported in part by the EU
Research Training Network {\it ForcesUniverse},
MRTN-CT-2004-005104.
\newpage

\appendix

\section{Some integral formulas}

In this section, we provide a complete list of integral formulas
needed for the computation of the correlation functions. As
reference we use \cite{abramowitz}.

We start with the formulas needed in the computation of boundary
three-point functions. First recall the integral representations
of the hypergeometric function $F(\alpha,\beta;\gamma|x)$
\begin{equation}
    \begin{split}
        &\int_1^\infty du \ |u|^{-\alpha}|u-1|^{-\beta}|u-x|^{-\gamma} \ = \\[2mm]
        &\qquad\qquad\qquad\qquad
        \frac{\Gamma(\alpha+\beta+\gamma-1)\Gamma(1-\beta)}{\Gamma(\alpha+\gamma)}F(\gamma,\alpha+\beta+\gamma-1;\alpha+\gamma\ | \ x) \\[2mm]
        &\int_0^x du \ |u|^{-\alpha}|u-1|^{-\beta}|u-x|^{-\gamma} \ = \\[2mm]
        &\qquad\qquad\qquad\qquad\qquad x^{1-\alpha-\gamma}\frac{\Gamma(1-\alpha)\Gamma(1-\gamma)}{\Gamma(2-\alpha-\gamma)}
        F(\beta,1-\alpha;2-\alpha-\gamma\ | \ x) \\[2mm]
        &\int_{-\infty}^0 du \ |u|^{-\alpha}|u-1|^{-\beta}|u-x|^{-\gamma} \ = \\[2mm]
%       &\int_1^\infty du \ |u|^{-\beta}|u-1|^{-\alpha}|u-(1-x)|^{-\gamma} \ = \\
        &\qquad\qquad\qquad\qquad\frac{\Gamma(\alpha+\beta+\gamma-1)\Gamma(1-\alpha)}{\Gamma(\beta+\gamma)}
        F(\gamma,\alpha+\beta+\gamma-1;\beta+\gamma\ | \ 1-x) \\[2mm]
        &\int_x^1 du \ |u|^{-\alpha}|u-1|^{-\beta}|u-x|^{-\gamma} \ = \\[2mm]
%       &\int_0^{1-x} du \ (-1)^{-\alpha-\gamma}|u|^{-\beta}|u-1|^{-\alpha}|u-(1-x)|^{-\gamma}\ = \\
        &\qquad\qquad\qquad\qquad(1-x)^{1-\beta-\gamma}\frac{\Gamma(1-\beta)\Gamma(1-\gamma)}{\Gamma(2-\beta-\gamma)}
        F(\alpha,1-\beta;2-\beta-\gamma\ | \ 1-x) \\[2mm]
    \end{split}
\end{equation}
these integrals converge for $|x|<1$.

If only the first order boundary interaction contributes, we need
the special case $\alpha+\beta+\gamma=2$ of the above integrals
which can be expressed as
\begin{equation}
    \begin{split}
        \int_{[-\infty,0]\ \cup\ [1,\infty]} du \ &|u|^{-\alpha}|u-1|^{-\beta}|u-x|^{-\gamma}  \ = \
%        (1-x)^{\alpha-1}\frac{\Gamma(1-\alpha)\Gamma(1-\beta)}{\Gamma(\gamma)}F(\alpha,1-\beta;\alpha\ | \ 1-x)\ =  \\
          (1-x)^{\alpha-1}x^{\beta-1}\frac{\Gamma(1-\alpha)\Gamma(1-\beta)}{\Gamma(\gamma)}\\[2mm]
\int_{[0,x]}\qquad\qquad  du \
&|u|^{-\alpha}|u-1|^{-\beta}|u-x|^{-\gamma} \ = \
          (1-x)^{\alpha-1} x^{\beta-1}\frac{\Gamma(1-\alpha)\Gamma(1-\gamma)}{\Gamma(\beta)}\\[2mm]
\int_{[x,1]}\qquad\qquad  du \
&|u|^{-\alpha}|u-1|^{-\beta}|u-x|^{-\gamma} \ = \
%       &\int_0^{1-x} du \ (-1)^{-\alpha-\gamma}|u|^{-\beta}|u-1|^{-\alpha}|u-(1-x)|^{-\gamma}\ = \
        (1-x)^{\alpha-1}x^{\beta-1}\frac{\Gamma(1-\beta)\Gamma(1-\gamma)}{\Gamma(\alpha)}\ .\\[2mm]
    \end{split}
\end{equation}

If the bulk interaction term contributes, we have to evaluate the
following integral for $\alpha+\beta+\gamma=0$
\begin{equation}
    \begin{split} \label{eq:A3}
        &\int d^2z \ \frac{(z-\bar{z})}{|z|^{2\alpha+2}|z-1|^{2\beta+2}|z-x|^{2\gamma+2}} \ = \\[2mm]
        &\ \ \ = \ \ \
    \frac{1}{\gamma x+\beta}\int d^2z\ \bar{\del}\ \Bigl(\frac{\bar{z}(\bar{z}-1)(\bar{z}-x)}{|z|^{2\alpha+2}|z-1|^{2\beta+2}|z-x|^{2\gamma+2}}\Bigr)\ +\\[2mm]
    &\ \ \ \ \ \ \ - \frac{1}{\gamma x+\beta}\int d^2z \ \del\ \Bigl(\frac{z(z-1)(z-x)}{|z|^{2\alpha+2}|z-1|^{2\beta+2}|z-x|^{2\gamma+2}}\Bigr) \\[2mm]
    &\ \ \ = \ -\frac{2}{\gamma x+\beta}\int du \ \frac{u(u-1)(u-x)}{|u|^{2\alpha+2}|u-1|^{2\beta+2}|u-x|^{2\gamma+2}} \\[2mm]
    &\ \ \ = \ -\frac{1}{\gamma(\gamma x+\beta)}\frac{d}{dx}\ \Bigl(\ \int_{[-\infty,0]\ \cup\ [1,\infty]} du \ \frac{1}{|u|^{2\alpha+1}|u-1|^{2\beta+1}|u-x|^{2\gamma}}\ +\\[2mm]
    &\ \ \ \ \ \ \ -\int_0^1 du \ \ \frac{1}{|u|^{2\alpha+1}|u-1|^{2\beta+1}|u-x|^{2\gamma}}\ \Bigr)\\[2mm]
        & \ \ \ = \ -4(1-x)^{2\alpha-1}x^{2\beta-1}\Bigl(\frac{\Gamma(-2\alpha)\Gamma(-2\beta)}{\Gamma(2\gamma+1)}+
        \frac{\Gamma(-2\alpha)\Gamma(-2\gamma)}{\Gamma(2\beta+1)}+\frac{\Gamma(-2\beta)\Gamma(-2\gamma)}{\Gamma(2\alpha+1)}\Bigr)\\[2mm]
    \end{split}
\end{equation}
and if two boundary interactions contribute, we need (again
$\alpha+\beta+\gamma=0$)
\begin{equation}
    \begin{split} \label{eq:A4}
      &\int_{a_1}^{b_1} du_1\ \int_{a_2}^{b_2}du_2 \ \frac{|u_1-u_2|}{|u_1u_2|^{\alpha+1}|(u_1-1)(u_2-1)|^{\beta+1}|(u_1-x)(u_2-x)|^{\gamma+1}} \ = \\[2mm]
      &\ \ \qquad\qquad = \ x^{2\beta-1}(1-x)^{2\alpha-1}\int_{c_1}^{d_1} du_1\
      \int_{c_2}^{d_2}du_2 \ \frac{|u_1-u_2|}{|(u_1-1)(u_2-1)|^{\beta+1}|u_1u_2|^{\gamma+1}} \ ,\\[2mm]
     \end{split}
 \end{equation}
 where $c_i=\frac{b_i^{-1}-x^{-1}}{1-x^{-1}}$ and $d_i=\frac{a_i^{-1}-x^{-1}}{1-x^{-1}}$. For these integrals one has to evaluate
 \begin{equation}
     \begin{split} \label{eq:A5}
         &\int_{1}^{\infty} du_1\
     \int_{1}^{u_1}du_2 \ \frac{(u_1-u_2)}{|(u_1-1)(u_2-1)|^{\beta+1}|u_1u_2|^{\gamma+1}} \ = \
     4\frac{\Gamma(-2\alpha)\Gamma(-2\beta)}{\Gamma(2\gamma+1)} \\[2mm]
         &\int_{0}^{1} du_1\
     \int_{0}^{u_1}du_2 \ \frac{(u_1-u_2)}{|(u_1-1)(u_2-1)|^{\beta+1}|u_1u_2|^{\gamma+1}} \ = \
     4\frac{\Gamma(-2\gamma)\Gamma(-2\beta)}{\Gamma(2\alpha+1)} \\[2mm]
         &\int_{-\infty}^{0} du_1\
     \int_{-\infty}^{u_1}du_2 \ \frac{(u_1-u_2)}{|(u_1-1)(u_2-1)|^{\beta+1}|u_1u_2|^{\gamma+1}} \ = \
     4\frac{\Gamma(-2\gamma)\Gamma(-2\alpha)}{\Gamma(2\beta+1)}  \\[2mm]
     \end{split}
\end{equation}
where we used the following special form of the Gamma doubling
formula
\begin{equation}
             \frac{\Gamma(1/2-\alpha)\Gamma(-\alpha)\Gamma(1/2-\beta)\Gamma(-\beta)}{\Gamma(1/2)\Gamma(\gamma+1/2)\Gamma(\gamma+1)} \ = \
         4\frac{\Gamma(-2\alpha)\Gamma(-2\beta)}{\Gamma(2\gamma+1)} \ .
\end{equation}

For the computation of bulk-boundary 2-point functions we use some
special cases of an integral formula that can be found in the recent
work of Fateev and Ribault \cite{Fateev:2007}. In case of a single
insertion of the bulk interaction we need
\begin{equation}
    \begin{split}\label{eq:1bulkint}
        \int d^2z\ \frac{|z-\bar{z}|}{|1+z^2|^{2(\alpha+1)}} \ = \ & -2i\pi^{3/2} 2^{-4\alpha}
        \frac{\Gamma(2\alpha+1/2)\Gamma(2\alpha)}{\Gamma^2(\alpha+1)\Gamma^2(\alpha+1/2)} \ .
    \end{split}
\end{equation}
To treat the insertion of one boundary interaction we employ
\begin{equation}
    \begin{split}\label{eq:1bdyint}
        \int du\ |1+u^2|^{-(\alpha+1)} \ = \ & \pi  2^{-2\alpha}\frac{\Gamma(2\alpha+1)}{\Gamma^2(\alpha+1)} \ .
    \end{split}
\end{equation}
The insertion of boundary interactions may be evaluated by means of the
following formula
\begin{equation}
    \begin{split}\label{eq:2bdyint}
        \int du_1du_2\ \frac{|u_1-u_2|}{|1+u_1^2|^{1+\alpha}|1+u_2^2|^{1+\alpha}} \ = \ & 4\pi^{3/2} 2^{-4\alpha}
        \frac{\Gamma(2\alpha+1/2)\Gamma(2\alpha)}{\Gamma^2(\alpha+1)\Gamma^2(\alpha+1/2)} \ .
    \end{split}
\end{equation}


\begin{thebibliography}{99}

%\cite{Schomerus:2005bf}
   \bibitem{Schomerus:2005bf}
       V.~Schomerus and H.~Saleur
     {\it The \GL\  WZW model: From supergeometry to logarithmic CFT},
     Nucl. Phys. {\bf B734} (2006) 221
     [arXiv:hep-th/0510032].
     %%CITATION = HEP-TH/0510032;%%"



%\cite{Rozansky:1992td}
\bibitem{Rozansky:1992td}
  L.~Rozansky and H.~Saleur,
  {\it S And T Matrices For The Super U(1$|$1) WZW Model: Application To Surgery And
  Three Manifolds Invariants Based On The Alexander-Conway Polynomial},
  Nucl.\ Phys.\  B {\bf 389} (1993) 365
  [arXiv:hep-th/9203069].
  %%CITATION = NUPHA,B389,365;%%

%\cite{Rozansky:1992zt}
\bibitem{Rozansky:1992zt}
  L.~Rozansky and H.~Saleur,
  {\it Reidemeister torsion, the Alexander polynomial and U(1$|$1) Chern-Simons
  Theory},
  J.\ Geom.\ Phys.\  {\bf 13} (1994) 105
  [arXiv:hep-th/9209073].
  %%CITATION = JGPHE,13,105;%%

%\cite{Flohr:2001zs}
\bibitem{Flohr:2001zs}
  M.~Flohr,
  {\it Bits and pieces in logarithmic conformal field theory},
  Int.\ J.\ Mod.\ Phys.\  A {\bf 18} (2003) 4497
  [arXiv:hep-th/0111228].
  %%CITATION = IMPAE,A18,4497;%%

%\cite{Gaberdiel:2001tr}
\bibitem{Gaberdiel:2001tr}
  M.~R.~Gaberdiel,
  {\it An algebraic approach to logarithmic conformal field theory},
  Int.\ J.\ Mod.\ Phys.\  A {\bf 18} (2003) 4593
  [arXiv:hep-th/0111260].
  %%CITATION = IMPAE,A18,4593;%%


  %\cite{Creutzig:2007}
  \bibitem{Creutzig:2007jy}
      T.~Creutzig, T.~Quella, and V.~Schomerus,
     {\it Branes in the \GL\  WZNW-Model},
     Nucl. Phys. {\bf B792} (2008) 257
     [arXiv:hep-th/0708.0583].
     %%CITATION = ARXIV:0708.0583;%%"

%\cite{Schomerus:1999ug}
\bibitem{Schomerus:1999ug}
  V.~Schomerus,
  {\it D-branes and deformation quantization},
  JHEP {\bf 9906}, 030 (1999)
  [arXiv:hep-th/9903205].
  %%CITATION = JHEPA,9906,030;%%

%\cite{Alekseev:1999bs}
\bibitem{Alekseev:1999bs}
  A.~Y.~Alekseev, A.~Recknagel and V.~Schomerus,
  {\it Non-commutative world-volume geometries: Branes on SU(2) and fuzzy
  spheres},
  JHEP {\bf 9909}, 023 (1999)
  [arXiv:hep-th/9908040].
  %%CITATION = JHEPA,9909,023;%%

%\cite{Alekseev:2002rj}
\bibitem{Alekseev:2002rj}
  A.~Y.~Alekseev, S.~Fredenhagen, T.~Quella and V.~Schomerus,
  {\it Non-commutative gauge theory of twisted D-branes},
  Nucl.\ Phys.\  B {\bf 646}, 127 (2002)
  [arXiv:hep-th/0205123].
  %%CITATION = NUPHA,B646,127;%%

%\cite{Schomerus:2002dc}
\bibitem{Schomerus:2002dc}
  V.~Schomerus,
  {\it Lectures on branes in curved backgrounds},
  Class.\ Quant.\ Grav.\  {\bf 19}, 5781 (2002)
  [arXiv:hep-th/0209241].
  %%CITATION = CQGRD,19,5781;%%

%\cite{Fateev:2007},
\bibitem{Fateev:2007}
    V.~Fateev and S.~Ribault,
     {\it Boundary action of the $H^+_3$ model},
     JHEP {\bf 02} (2008) 024
     [arXiv:hep-th/0710.2093].
     %%CITATION = ARXIV:0710.2093;%%"

%\cite{Warner:1995ay}
\bibitem{Warner:1995ay}
  N.~P.~Warner,
  {\it Supersymmetry in boundary integrable models},
  Nucl.\ Phys.\  B {\bf 450} (1995) 663
  [arXiv:hep-th/9506064].
  %%CITATION = NUPHA,B450,663;%%

%\cite{Kapustin:2003ga}
\bibitem{Kapustin:2003ga}
  A.~Kapustin and Y.~Li,
  {\it Topological correlators in Landau-Ginzburg models with
  boundaries},
  Adv.\ Theor.\ Math.\ Phys.\  {\bf 7} (2004) 727
  [arXiv:hep-th/0305136].
  %%CITATION = 00203,7,727;%%

%\cite{Brunner:2003dc}
\bibitem{Brunner:2003dc}
  I.~Brunner, M.~Herbst, W.~Lerche and B.~Scheuner,
  {\it Landau-Ginzburg realization of open string TFT},
  JHEP {\bf 0611} (2006) 043
  [arXiv:hep-th/0305133].
  %%CITATION = JHEPA,0611,043;%%

%\cite{Hosomichi:2004ph}
\bibitem{Hosomichi:2004ph}
  K.~Hosomichi,
  {\it $N = 2$ Liouville theory with boundary},
  JHEP {\bf 0612} (2006) 061
  [arXiv:hep-th/0408172].
  %%CITATION = JHEPA,0612,061;%%

  %{Gerasimov:1990}
  \bibitem{Gerasimov:1990}
     A.~Gerasimov, A.~Morozov, M.~Olshanetsky, A.~Marshakov and S.~L.~Shatashvili,
     {\it Wess-Zumino-Witten model as a theory of free fields},
     Int.\ J.\ Mod.\ Phys.\ {\bf A5} (1990) 2495
     %%CITATION = IMPAE,A5,2495;%%"

%\cite{Gotz:2006qp}
\bibitem{Gotz:2006qp}
  G.~G\"otz, T.~Quella and V.~Schomerus,
  {\it The WZNW model on PSU(1,1$|$2)},
  JHEP {\bf 0703} (2007) 003
  [arXiv:hep-th/0610070].
  %%CITATION = JHEPA,0703,003;%%

%\cite{Gaberdiel:2006pp}
\bibitem{Gaberdiel:2006pp}
  M.~R.~Gaberdiel and I.~Runkel,
  {\it The logarithmic triplet theory with boundary},
  J.\ Phys.\ A  {\bf 39} (2006) 14745
  [arXiv:hep-th/0608184].
  %%CITATION = JPAGB,A39,14745;%%

%\cite{LeClair:2007aj}
\bibitem{LeClair:2007aj}
  A.~LeClair,
  {\it The gl(1$|$1) super-current algebra: The role of twist and logarithmic
  fields},
  arXiv:0710.2906 [hep-th].
  %%CITATION = ARXIV:0710.2906;%%


     %\cite{Creutzig:2006wk}
     \bibitem{Creutzig:2006}
     T.~Creutzig, T.~Quella, and V.~Schomerus,
     {\it New boundary conditions for the $c = -2$ ghost system},
     Phys. Rev. {\bf D77} (2008) 026003
     [arXiv:hep-th/0612040].
     %%CITATION = HEP-TH/0612040;%%"





%\cite{Quella:2007hr}
\bibitem{Quella:2007hr}
  T.~Quella and V.~Schomerus,
  {\it Free fermion resolution of supergroup WZNW models},
  JHEP {\bf 0709} (2007) 085
  [arXiv:0706.0744 [hep-th]].
  %%CITATION = JHEPA,0709,085;%%


%\cite{Frappat:1996pb}
\bibitem{Frappat:1996pb}
  L.~Frappat, P.~Sorba and A.~Sciarrino,
  {\it Dictionary on Lie superalgebras},
  arXiv:hep-th/9607161.
  %%CITATION = HEP-TH/9607161;%%

\bibitem{abramowitz}
    M.~Abramowitz and I.~A.~Stegun
    {\it Handbook of mathematical functions with formulas,
    graphs, and mathematical tables},
    National Bureau of Standards Applied Mathematics Series, 55


\end{thebibliography}
\end{document}